\documentclass[twocolumn]{aastex63}

\received{June 1, 2019}
\revised{January 10, 2019}
\accepted{\today}

\submitjournal{ApJ}

\shorttitle{S62 and S4711}
\shortauthors{Pei$\beta$ker et al.}

\graphicspath{{./}{figures/}}

\begin{document}

\title{S62 and S4711: Indications of a population of faint fast moving stars inside the S2 orbit\\
       \small S4711 on a 7.6 year orbit around Sgr~A*}

\correspondingauthor{Florian Pei{\ss}ker}
\email{peissker@ph1.uni-koeln.de}
\author[0000-0002-9850-2708]{Florian Pei$\beta$ker}
\affil{I.Physikalisches Institut der Universit\"at zu K\"oln, Z\"ulpicher Str. 77, 50937 K\"oln, Germany}
\author[0000-0001-6049-3132]{Andreas Eckart}
\affil{I.Physikalisches Institut der Universit\"at zu K\"oln, Z\"ulpicher Str. 77, 50937 K\"oln, Germany}
\affil{Max-Plank-Institut f\"ur Radioastronomie, Auf dem H\"ugel 69, 53121 Bonn, Germany}

\author[0000-0001-6450-1187]{Michal Zaja\v{c}ek}
\affil{Center for Theoretical Physics, Al. Lotników 32/46, 02-668 Warsaw, Poland}
\affil{I.Physikalisches Institut der Universit\"at zu K\"oln, Z\"ulpicher Str. 77, 50937 K\"oln, Germany}

\author{Basel Ali}
\affil{I.Physikalisches Institut der Universit\"at zu K\"oln, Z\"ulpicher Str. 77, 50937 K\"oln, Germany}

\author{Marzieh Parsa}
\affil{I.Physikalisches Institut der Universit\"at zu K\"oln, Z\"ulpicher Str. 77, 50937 K\"oln, Germany}

\begin{abstract}

We present high-pass filtered NACO and SINFONI images of the newly discovered stars S4711-S4715 between 2004 and 2016.
Our deep H+K-band (SINFONI) and K-band (NACO) data shows the S-cluster star S4711 on a highly eccentric trajectory around Sgr~A* with an orbital period of 7.6 years and a periapse distance of 144 AU to the super massive black hole (SMBH).
S4711 is hereby the star with the shortest orbital period and the smallest mean distance to the
SMBH during its orbit to date.
The used high-pass filtered images are based on co-added data sets to improve the signal to noise.
The spectroscopic SINFONI data let us determine detailed stellar properties of S4711 like the mass and the rotational velocity. 
The faint S-cluster star candidates, S4712-S4715, can be observed in a projected distance to Sgr~A* of at least temporarily $\leq$ 120 mas. From these stars, S4714 is the most prominent one with an orbital period of 12 years and an eccentricity of 0.985.
The stars S4712-S4715 show similar properties with comparable magnitudes and stellar masses to S4711.
The MCMC simulations determine confidently precise uncertainties for the orbital elements of S62 and S4711-S4715.
The presence of S4711 in addition to S55, S62, and the also newly found star S4714 implies a population of faint stars that can be found at distances to Sgr~A* that are comparable to the size of our solar system.
These short orbital time period stars in the dense cluster around the SMBH in the center of our Galaxy are perfect candidates to observe gravitational effects such as the periapse shift.

\end{abstract}



\section{Introduction}

Since the first detection of stellar proper motions around the super massive black hole (SMBH) Sagittarius A* (Sgr~A*) by \citet{Eckart1996} and \citet{Eckart1997}, the number of detected S-stars increased with the development of larger telescopes and advanced analysis techniques \citep{Ghez1998,Ghez2002,Gillessen2009,2010RvMP...82.3121G,Gillessen2017,gravity2018,Gravitycollaboration2020}.
Recently, we presented the orbit of S62 \citep{peissker2020a} as obtained with the Very Large Telescope (VLT, Chile). This star moves on a highly eccentric trajectory of $e\,=\,0.976$ around Sgr~A* with an orbital time period of $t_{\rm period}\,=\,9.90\,\pm\,0.02$ years. The distance of S62 to Sgr~A* is $r_{\rm p} \,=\,16.0\,$AU based on the well defined orbital elements. The detection of this S-cluster member is in line with the observation of S55 \citep[also known as S0-102, see][]{MeyerS552012}. S55 has an orbital timescale of $t_{\rm period}\,=\, 12.8\,\pm\,0.1$ years with an eccentricity of $e\,=\,0.7209$ and has been detected with the KECKII telescope (Hawaii). The highly eccentric orbits of these faint stars (compared to $m_K\,=\,14.1$ for S2) raise the question of their origin and their uniqueness. 

Stars in the central parsec of the Galactic Center and in particular the S-cluster members cannot have formed under classical conditions. Considering the presence of a high mass concentration and the resulting tidal forces excludes an in situ star formation scenario. This is also summarized in the phrase ``Paradox of youth" formulated by \cite{Ghez2003}.
While \cite{Nayakshin2007} and \cite{Jalali2014} show how stars could have formed in the central parsec or even in the immediate vicinity of the SMBH, there remains the
question of how stars can be placed on highly elliptical orbits with such small separations from Sgr~A*.
The Hills mechanism provides a matching theoretical background to describe the situation \citep{Hills1988}. The Hills mechanism describes a three-body interaction including Sgr~A* black hole and an eccentric stellar binary. When the binary (star-A and star-B) approaches Sgr~A*, it gets disrupted and one companion (star-A) moves on a circularized orbit while the star-B gains the energy at the expanse of star-A and can be eventually expelled as a runaway star on an escaping hyperbolic orbit. This runaway star (star-B) could be kicked out of the binary system with hypervelocity ($>$ 300 km/s). Another possible product of this interaction could be two single stars on highly-eccentric orbits, one with a smaller eccentricity and the other with a larger eccentricity \citep[see numerical simulations done by several authors, ][]{2003ApJ...592..935G,2007ApJ...656..709P,2008ApJ...683L.151L,2012ApJ...749L..42B,Zajacek2014}.

The three-body interaction including the massive black hole can effectively fill the `sparse region' around Sgr~A*. In the classical quasi-spherical stellar cluster around Sgr~A* with the power-law distribution of stellar number densities, there is a well-defined radius below which the number of stars statistically drops below one \citep{2018AN....339..324Z},
\begin{equation}
    r_{1}=r_{\rm h} N_{\rm h}^{-\frac{1}{3-\gamma}}\,,
    \label{eq_radius_1}
\end{equation}
where $r_{\rm h}$ is the gravitational influence radius of Sgr~A* ($r_{\rm h}\sim 2\,{\rm pc}$), $N_{\rm h}$ is the total number of stars inside $r_{\rm h}$ and $\gamma$ is the slope of the stellar number density distribution, $n_{\star} \propto r^{-\gamma}$.
For main-sequence stars, $r_1$ is expected to be at $\sim 500$ Schwarzschild radii \citep[or about 40 AU, ][]{2006ApJ...645.1152H}. The Hills mechanism and scattering by massive perturbers \citep{2007ApJ...656..709P} can effectively fill this `sparse region' with stars on orbits with large eccentricities and pericenter distances of $r_{\rm p} \lesssim r_1$.

The S-stars S2, S55, and S62 are already three stars that are orbiting Sgr~A* with orbital parameters that can be described with a high eccentricity and a small semi-major axis of a few milliparsecs. Hence, these stars could be dynamical remainders of the Hills mechanism. Two of them, namely S55 and S62, orbit Sgr~A* with an orbital period of the order of 10 years, which makes them convenient probes of general relativistic effects. This is especially encouraging as there might be even more stars with comparable orbital elements filling the `sparse region' around Sgr~A* as already predicted by \cite{Alexander2003a} where the authors introduce the existence of the so-called `squezzars'. \cite{Alexander2003a} predict around $120$ stars with $r_p \,<\,120.0\,AU$ and $t_{\rm period}\,<\, 60$ years in the Galactic center.

In this work, we present the high eccentric orbit of a new star inside the orbit of S2. We cover a full orbit of the new S-cluster member with the detection in the high-pass filtered NACO K-band images. The data reduction and the applied methods are described in Sec. \ref{sec:data}. In Sec. \ref{sec:results}, we show the results of the orbital fit and the simultaneous observation of S4711 in the SINFONI and NACO data. From the SINFONI data, we derive a Doppler shifted line of sight (LOS) velocity. Additionally, we present Markov Chain Monte Carlo (MCMC) simulations of the orbital elements. From these simulations, we derive the uncertainties for the well-defined orbital elements of S4711 and S62. In Sec. \ref{sec:discussion}, we discuss the results and conclude the here presented analysis.

\section{Data \& Observations}
\label{sec:data}
In this section, we will briefly describe the analyzing methods and the data reduction steps. Most of the analyzing steps are presented and based on \cite{Parsa2017}, \cite{peissker2020a}, and \cite{peissker2020b}. It should be noted, that the authors of \cite{Parsa2017} analysed the orbit of the S-star S2 (also know as S0-2) and showed the relativistic Schwarzschild precession with this very NACO data-set. This was later confirmed by \cite{Gravitycollaboration2020} and underlines the robustness of the data-set.

\subsection{Data reduction}

For this work, we used the SINFONI and NACO data sets that are already discussed, shown, and analysed in \citep{Peissker2019, peissker2020a, peissker2020b}. For the K-band observations, we use the NAOS+CONICA (NACO) instrument with the adaptive optics (AO) guide star IRS7 located about 5".5 north of Sgr~A*. The H+K observations are carried out with the Spectrograph for Integral Field Observations in the Near Infrared (SINFONI). The optical guide star for the SINFONI AO correction is located 15”.54 north and 8”.85 east of Sgr~A*. Both instruments were mounted at the Very Large Telescope (VLT) in Paranal/Chile and are now decommissioned. Standard data reduction steps with the ESO pipeline were applied. For the SINFONI H+K data, we use DARK frames to correct for hot pixels. LINEARITY frames are taken to supervise the detector response. FLAT lamps measure the pixel-to-pixel response. DISTORTION frames are used to correct for optical distortions but also to monitor the slitlet distances. Lastly, WAVE frames are taken in order to wavelength calibrate the data. Since we also use NACO in the K-band imaging mode, the reduction steps are comparably short. We apply the FLAT field and bad pixel correction. For creating a mosaic, we shift the individual images in an 2048$\times$2048 array to their related positions. Furthermore, the data from both instruments have been sky corrected. Tables of the used data are listed in Sec. \ref{sec_app:data_tables}, Appendix.

\subsection{Methods}

For reducing the influence of overlapping Point Spread Function (PSF) wings, a low pass filter is a suitable and efficient approach. We are subtracting a Gaussian flux density conserved smoothed image from the input image. For that, the size of the Gaussian filter should be close to the PSF size. In SINFONI, this is usually around 5-6 pixels corresponding to 62-75 mas and depends on the overall data-quality. After the subtraction, we smooth the result again. In principal, the same Gaussian filter size is possible. However, we use different filter sizes of less than 6 pixel. The result that
allows the best discrimination of densely packed stars from each other is then chosen manually. Since there is no loss in information, the process is stable against false positives \cite[see also][]{peissker2020b}.\newline
Another method for analysing the data is the high pass filter, also known as the Lucy Richardson algorithm \citep{Lucy1974}. Since noise can be associated with low frequencies, the high pass filter helps to highlight image details that are above a self defined threshold. It is known, that the Galactic center and especially the S-cluster suffers from a spatially variable background. This leads for example to the creation of blended stars \citep{Sabha2012}. With this, a background subtraction is sufficient and needs to be adjusted for every data set because of different weather conditions and the varying number of high quality observations. Cosmic rays, the influence of the detector cosmetics, read-out errors and the overall state of the data are noticeable error sources. 
The determination of the position of Sgr~A* is based on the well known and observed orbit of S2 \citep{Parsa2017, Gravity2019}.

\section{Results}
\label{sec:results}
Here, we present our results of the SINFONI and NACO data analysis. We show the orbit of S4711 and the detection of the star in the NACO and SINFONI data. From the latter one, we extract a Doppler shifted spectrum and derive a LOS velocity for S4711. Furthermore, we present the observation of S4712-S4715. Some of these S-stars are at least temporarily close to Sgr~A* with comparable properties to those of S4711. Complementary material can be found in the Appendix. 

\subsection{S4711 on a 7.6 yr orbit around Sgr~A*}

\begin{figure}[ht!]
	\centering
	\includegraphics[width=.5\textwidth]{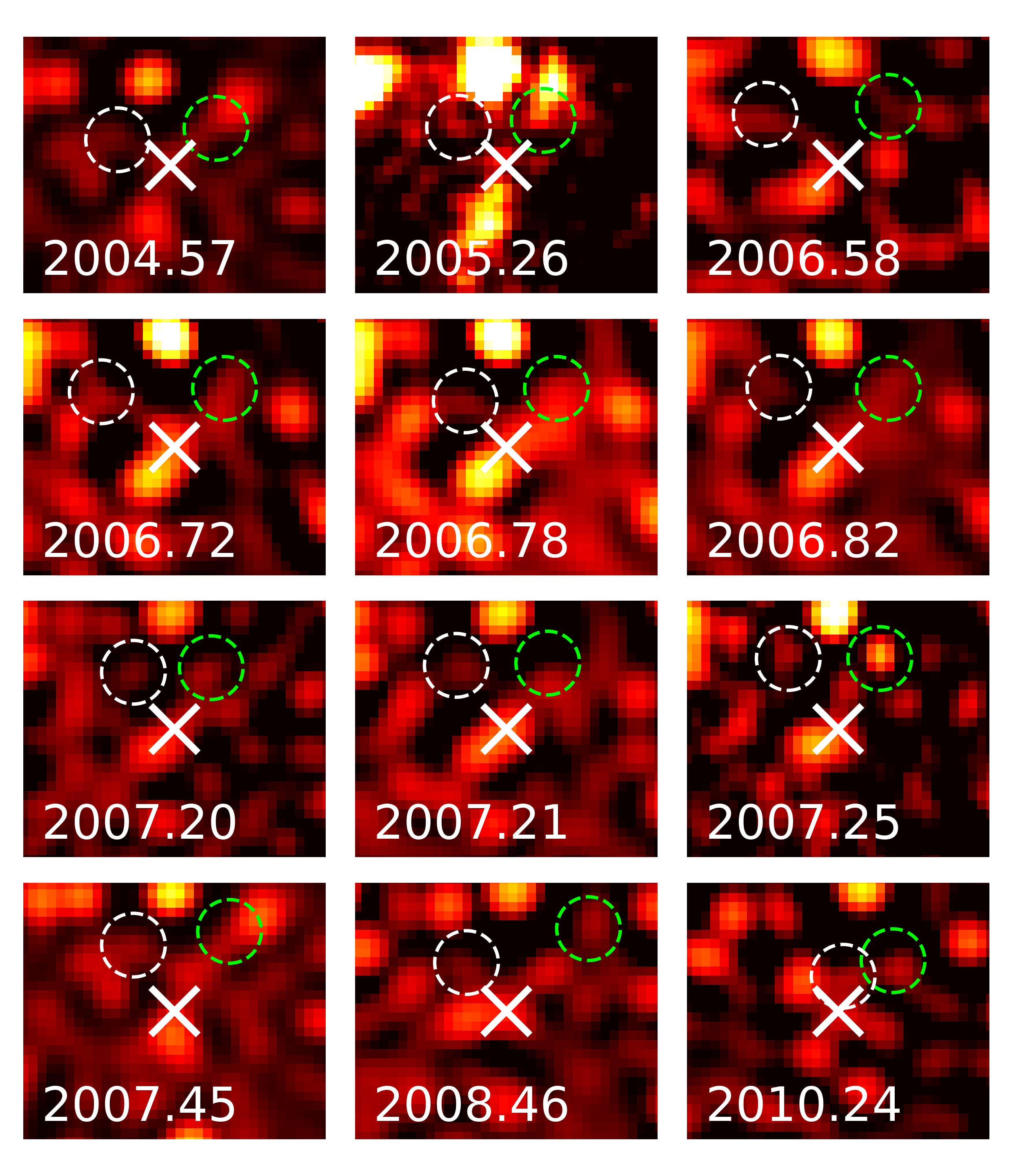}
	\caption{S4711 on its orbit around Sgr~A*. Here, we present NACO K-band images that are high pass filtered. Sgr~A* is marked with a white cross and the position of S62 is marked with a green dashed circle. S4711 is highlighted by a white dashed circle. Orientation-wise, North is up and east is to the left. The size of every image in this figure is about 0".44$\times$0".37.}
\label{fig:orbit_timeline}
\end{figure}

\begin{figure*}[htbp!]
	\centering
	\includegraphics[width=1.\textwidth]{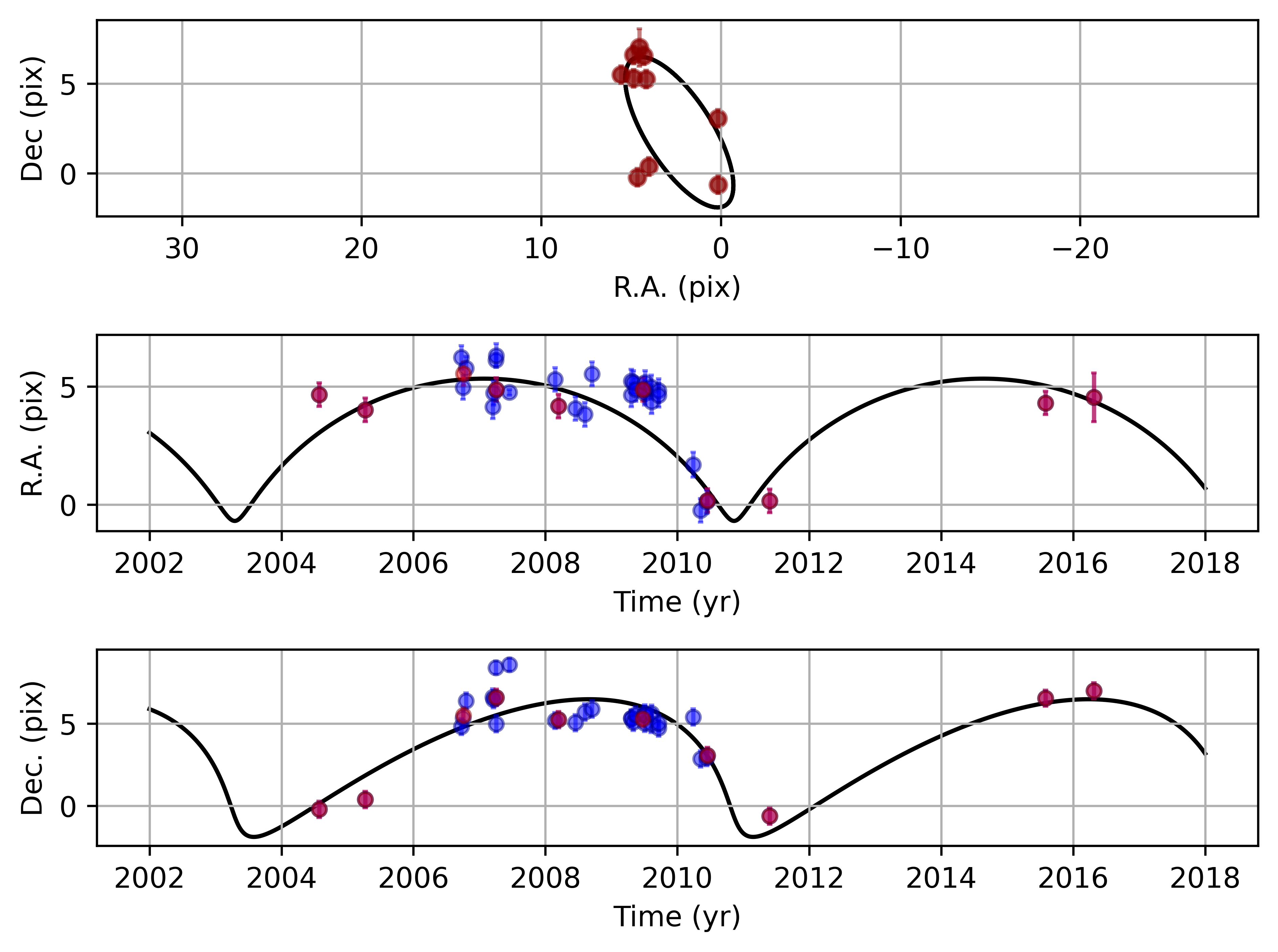}
	\caption{Orbit of S4711 where 1 pixel corresponds to 13.3 mas. The plot shows all available data points with the typical uncertainties of $\pm\,6.5$ mas based on the NACO data. Red data points represent the geometric median of most of the years except for 2004, 2005, 2011, 2015, and 2016. The blue data points show the data. Because of S17, S62 and the DSO/G2 object, an identification of S4711 between 2012 and 2013 is not possible without confusion. In 2014, NACO was not mounted at the VLT. In 2017 and 2018, the periapse passage of S2 hinders a successful observation of S4711.}
\label{fig:orbit}
\end{figure*}

\begin{figure}[htbp!]
	\centering
	\includegraphics[width=.5\textwidth]{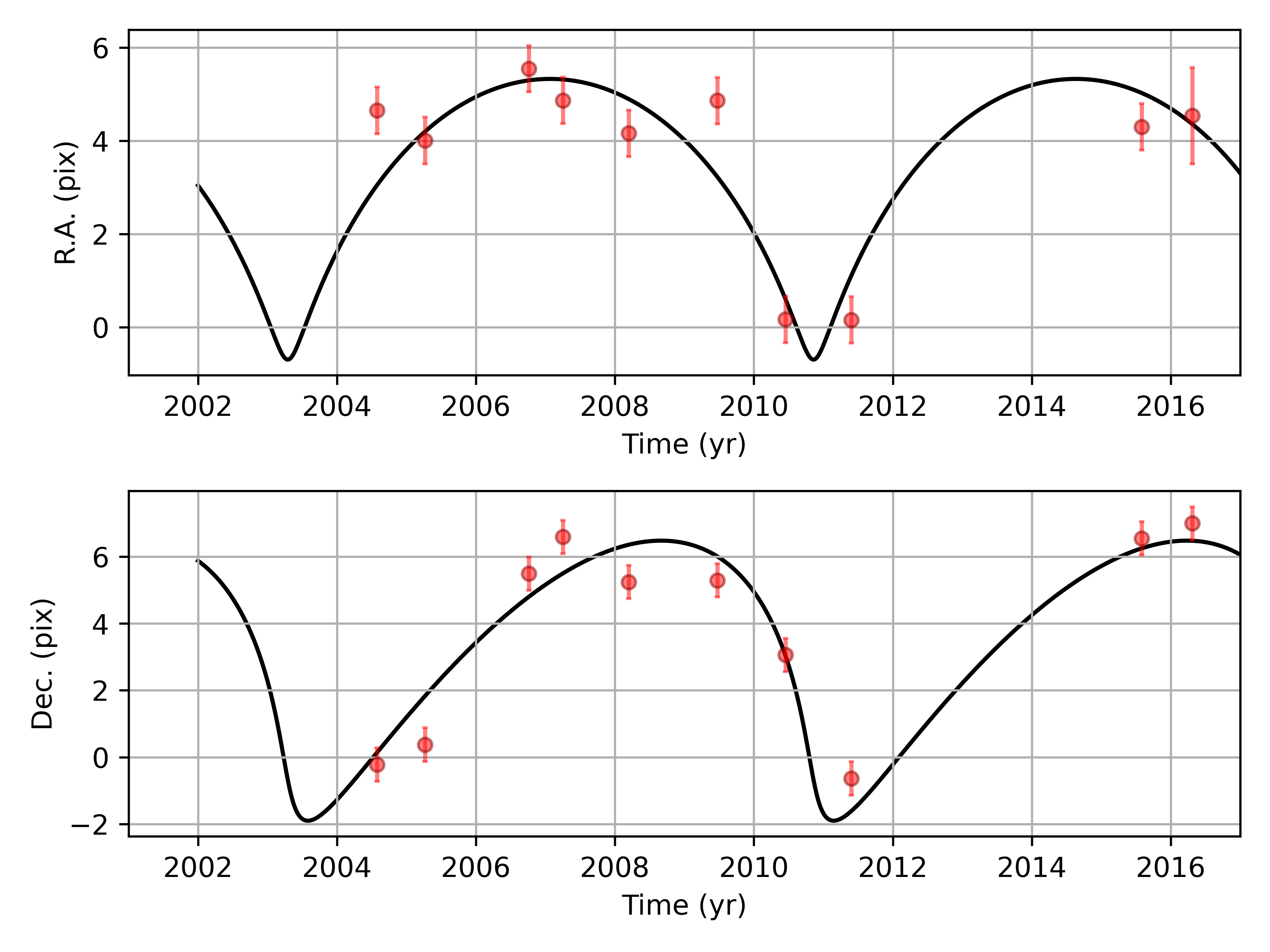}
	\caption{Median orbit of S4711. Here, we show the median of every year. The median is stable against the influence of outliers. 1 pixel corresponds to 13.3 mas. Similar to Fig. \ref{fig:orbit_timeline}, we set the units of the y-axis to [pixel] to underline the close distance to Sgr~A*.}
\label{fig:orbit_median}
\end{figure}

The close passage of S2 did not allow sensitive observations of the
region covered by the S4711 orbit in 2017 until 2019.
However, for every year during the time span from 2004 to 2016, we can confirm several detections of S4711 in the high-pass filtered NACO K-band images (Fig. \ref{fig:orbit_timeline} and Fig. \ref{fig:orbit}).
Source positions are derived via Gaussian fits to the objects in the images of each epoch.
Since the positional determinations in the high resolution images are not necessarily Gaussian 
distributed, we used the median in the years between 2006-2010 to avoid outliers (Fig. \ref{fig:orbit_median}). Consider also the detailed description of the analysis given in \cite{peissker2020a}. Based on the positions 
derived from these data-sets we were able to fit the orbit of S4711 (indicated by red data-points in Fig.\ref{fig:orbit}). The resulting orbital elements are presented in Table \ref{tab:orbital_para}.\newline
The maximum distance of S4711 from Sgr~A* is around 80 mas ($\approx 6$ pixel in the NACO images). Therefore, the star suffers blending and confusion with other objects in the crowded field. This effect is minimized because of the constant monitoring of the S-cluster in the last 25 years but underlines the need for a sufficient number of high-quality observations.
\begin{table*}[htbp!]
    \centering
    \begin{tabular}{cccccccc}
            \hline
            \hline
            Source & a [mpc] & e & i [$^\circ$] & $\omega$ [$^\circ$] & $\Omega$ [$^\circ$]&  $t_{\rm closest}$ [years]& $t_{\rm period}$ [years] \\
            \hline
             S62   & 3.588 $\pm$ 0.02 &  0.976 $\pm$ 0.01 & 72.76 $\pm$ 5.15 & 42.62 $\pm$ 2.29 & 122.61 $\pm$ 4.01 & 2003.33 $\pm$ 0.02 & 9.9 $\pm$ 0.3 \\ 
             S4711   & 3.002 $\pm$ 0.06  &  0.768 $\pm$ 0.030  & 114.71 $\pm$ 2.92 & 131.59 $\pm$ 3.09 & 20.10 $\pm$ 3.72 & 2010.85 $\pm$ 0.06 & 7.6 $\pm$ 0.3 \\
            S4712  & 18.038 $\pm$ 0.099 &  0.364 $\pm$ 0.032 & 117.28 $\pm$ 1.31 & 238.08 $\pm$ 3.43 & 166.38 $\pm$ 3.20 & 2007.12 $\pm$ 0.08 & 112.0 $\pm$ 2.9 \\   
             S4713   & 8.016 $\pm$ 0.379 &  0.351 $\pm$ 0.059 & 111.07 $\pm$ 1.66 & 301.97 $\pm$ 8.02 & 195.06 $\pm$ 5.15 & 2000.03 $\pm$ 0.22 &  33.2 $\pm$ 2.5 \\    
             S4714 & 4.079 $\pm$ 0.012 & 0.985 $\pm$ 0.011 & 127.70 $\pm$ 0.28 & 357.25 $\pm$ 0.80 & 129.28 $\pm$ 0.63 & 2017.29 $\pm$ 0.02 & 12.0 $\pm$ 0.3 \\
             S4715 & 5.756 $\pm$ 0.439 & 0.247 $\pm$ 0.040 & 129.80 $\pm$ 3.72 & 359.99 $\pm$ 5.38 & 282.15 $\pm$ 2.92 & 2008.05 $\pm$ 0.30 & 20.2 $\pm$ 2.4\\
            \hline
    \end{tabular}
    \caption{Orbital parameters for S4711, S4712, S4713, S4714, S4715, and S62. The 1$\sigma$ uncertainty is based on the MCMC simulations. The orbital period was calculated assuming the Sgr~A* mass of $(4.1 \pm 0.2) \times 10^6\,M_{\odot}$. The related uncertainty was determined by the error propagation.}
    \label{tab:orbital_para}
\end{table*}

\begin{figure*}[htbp!]
	\centering
	\includegraphics[width=.8\textwidth]{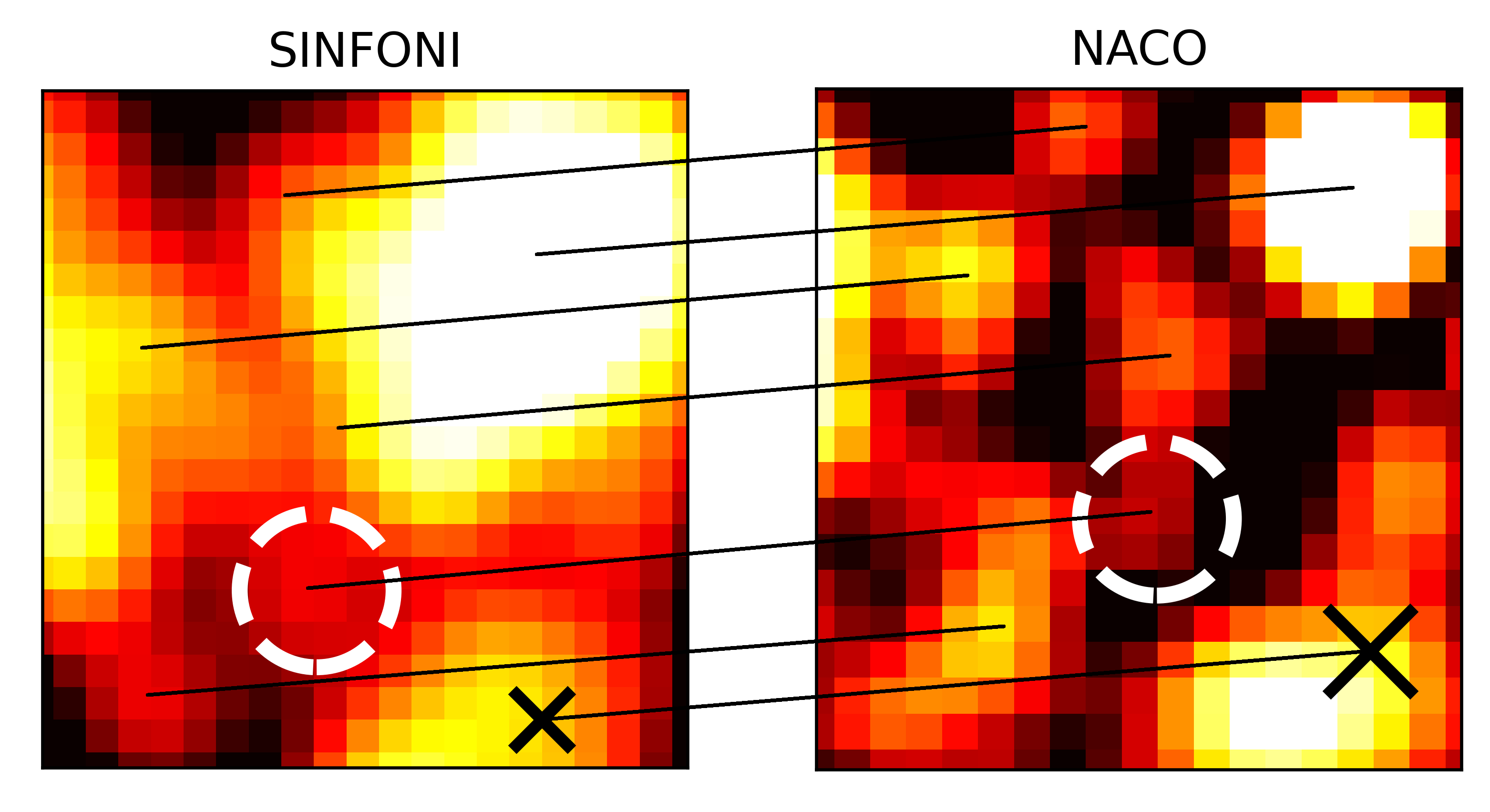}
	\caption{SINFONI and NACO detection of S4711 in 2007 based on high pass filtered images. The S-star S2 is located in the upper right corner. Sgr~A* is marked with a black cross. The white dashed circle indicates the position of S4711. The position of the stellar sources in the SINFONI and NACO data is consistent with each other. For comparison, we indicate the detection and position in both data sets with a black line. Based on the orbital fit, the chance of detecting S4711 is maximized because of its isolated position in 2007. 10000 iterations are used to neglect the chance of a false positive. We use an artificial PSF with similar properties as a natural one. Because of the contrast and the subtracted background, some areas seem to be under-crowded compared to others. Since both instruments have a different plate scale, distances appear distorted. In every image, north is up, east is to the left.}
\label{fig:sinfo_2007}
\end{figure*}
\begin{figure}[htbp!]
	\centering
	\includegraphics[width=.5\textwidth]{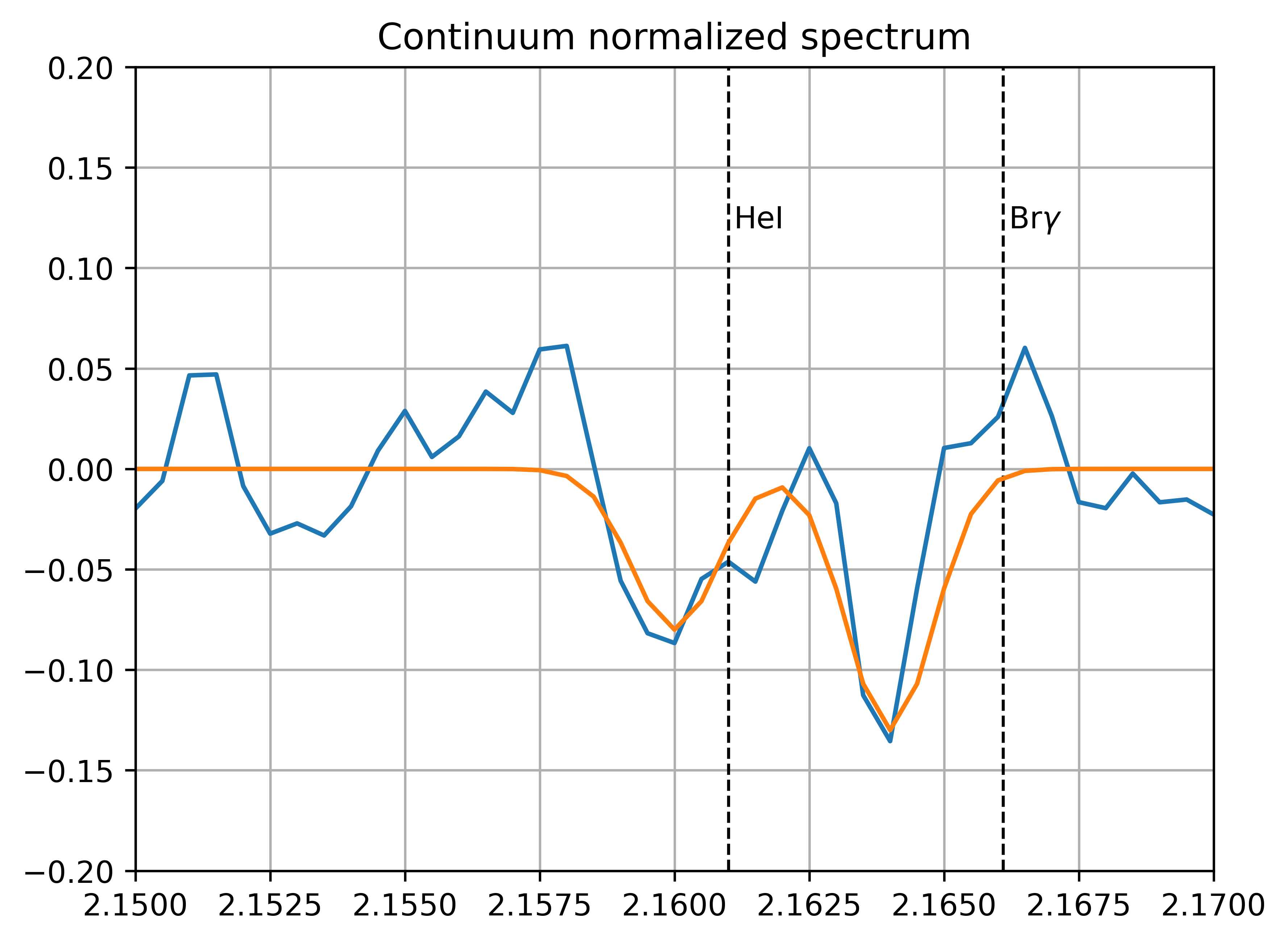}
	\caption{Normalized spectrum of S4711 extracted from the SINFONI data of 2007. We use an iris-aperture with $r_{\rm in}\,=\,1$px and $r_{\rm out}\, =\, 3$px and refer for the selected aperture to Fig. \ref{fig:sinfo_2007}, left plot. We find both a Doppler-shifted He$I$ and Br$\gamma$ absorption lines with related LOS velocity of $v_{\rm Br\gamma}\,=\,-291\,\pm\,55$km/s \citep[for comparable values, see][]{Habibi2017}.}
\label{fig:sinfo_2007_spec}
\end{figure}

Based on the NACO and SINFONI detection of S4711 (see Fig. \ref{fig:orbit_timeline} and Fig. \ref{fig:sinfo_2007}), we find a K-band magnitude of $m_K\,=\,18.43\,\pm\,0.22$ with
\begin{equation}
    m_{\rm K, S4711} = -m_{\rm K, S2} + 2.5\times \log(ratio)
\end{equation}
with aperture photometry and a S2 K-band magnitude of 14.15 \citep{Schoedel2010}. We use this method to derive the apparent magnitude m$_K$ for all newly discovered stars (see Table \ref{tab:stellar_para}). To determine the mass of these faint S-cluster stars (FSS), we adapt from \cite{Cai2018}
\begin{equation}
    lg\frac{M_{\rm S}}{M_\odot}\,=\,km_K+b
    \label{eq:mass}
\end{equation}
with $k\,=\,-0.192$ and $b\,=\,3.885$. The authors of \cite{Cai2018} derive from a fit of the S-star properties presented in \cite{Habibi2017} the values for k and b. Using the same data from \cite{Habibi2017}, we confirm $b\,=\,3.885$ and use in the following $k\,=\,-0.1925$. Using Eq. \ref{eq:mass} with the values given in \cite{Habibi2017}, we find an agreement with the stellar mass in the range of $5\%\,-\,10\%$ for the there discussed S-stars\footnote{S1, S2, S4, S6, S7, S8, S9, and S12}. We adapt an error of $-\,1\,M_\odot$ and $+\,2\,M_\odot$. With that, Equation \ref{eq:mass} provides an accessible approach for deriving the mass for the S-stars when no spectroscopic information is available. Consult Table \ref{tab:stellar_para} for the stellar mass of S62 and S4711-S4715.\newline\newline
\begin{table}[htbp!]
    \centering
    \begin{tabular}{ccc}
            \hline
            \hline
            Source & Magnitude [mag$_K$] & Mass [M$_{\odot}$] \\
            \hline
             S62    & 16.1 & 6.1 \\
             S4711  & 18.4 & 2.2 \\
             S4712  & 18.4 & 2.2 \\   
             S4713  & 18.5 & 2.1 \\    
             S4714  & 17.7 & 2.0 \\
             S4715  & 17.8 & 2.8 \\
             
            \hline
    \end{tabular}
    \caption{Stellar parameters for known and newly discovered S-stars. For the photometric results of S62 and S4711-S4715, we consistently use the S-cluster member S2 as a reference star. We use the Eq. \ref{eq:mass} to derive the mass of these S-cluster members. The mass of S62 is an updated value \citep[see][]{peissker2020a}. Typical uncertainties for the magnitude are $\pm\,0.2$, the stellar mass is in the range of $-\,1\,M_\odot$ and $+\,2\,M_\odot$ \citep{Habibi2017, Cai2018, peissker2020a}.}
    \label{tab:stellar_para}
\end{table}
In 2007, S4711 reaches it apoapse. Also, the distance between S4711 and S2 is sufficient enough to analyse the spectrum of the newly discovered star in the SINFONI data of the related year (Fig. \ref{fig:sinfo_2007_spec}). We find a line of sight (LOS) velocity of $v_{\rm Br\gamma}\,=\,-291\,\pm\,55$km/s. The depth of the HeI and Br$\gamma$\footnote{Rest wavelength: HeI at $2.161\mu m$, Br$\gamma$ at $2.166\mu m$} absorption features are comparable with the analysis of the S-cluster stars presented in \cite{Habibi2017}. Because of the apoapse position of S4711, the derived LOS velocity is a lower limit of the range of possible Dopplershifted values.
The observed HeI line at $2.161\mu m$ is actually a doublet, namely $7^{3}F-4^{3}D$ at $2.16137\,\mu m$ and $7^{1}F-4^{1}D$ at $2.16229\,\mu m$ \citep{Lumsden2001}. This "shoulder" is clearly detectable in the spectrum (Fig. \ref{fig:sinfo_2007_spec}) and shows an asymmetrical shape. This points towards a low rotational velocity ($v\,\sin{i}$) of the star. A singlet would be the result of a high $v\,\sin{i}$ value. The presence of a rotational velocity, however, prolongs the lifetime of the star \citep{Clark2018} and therefore possible in-spiral events that will be discussed in Sec.\ref{sec:discussion}. Considering the critical study of the rotational velocity by \cite{Zorec2017}, we determine for S4711 $v\,\sin{i}\,=\,239.60\,\pm\,25.21\,km/s$ with the FWHM of the $2.161\mu m$ HeI line. Even though \cite{Slettebak1975} finds somewhat lower values for the Br$\gamma$ line, the here derived rotational velocity is in a comparable range with the findings of \cite{Clark2000} and \cite{Abt2002}.

\begin{figure}[htbp!]
	\centering
	\includegraphics[width=.5\textwidth]{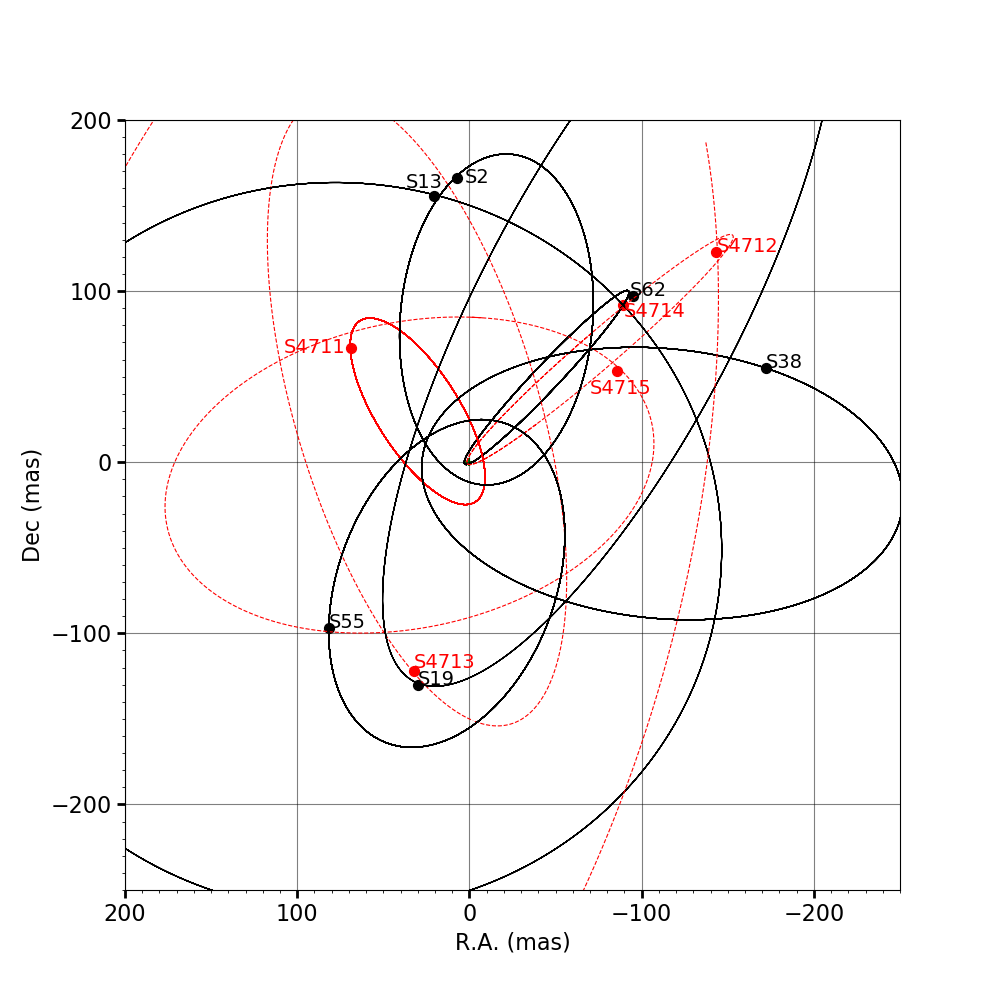}
	\caption{Orbits of several S-stars close to Sgr~A*. The SMBH is located at the origin of the reference frame. The orbit of the here reported S-cluster star S4711 is highlighted with solid red, as for S4712, S4713, S4714 and S4715, their orbits are marked with dashed red. The dots represent the corresponding stellar position in 2007.}
\label{fig:all_orbits}
\end{figure}

Furthermore, we derive from the shape of the Br$\gamma$ line the spectral type of S4711. We use the NIR spectral atlas of \cite{Hanson1996} and follow the analysis presented in \cite{Eisenhauer2005} to determine the spectral type of S4711. For the Br$\gamma$ line width ($\sim 10\,\mathring{A}$) and the spectrum, we find a good agreement with the emission of a B8/9-V star with an effective temperature of about $T_{\rm eff}\,=\,10700\,-\,11800$K \citep{Tokunaga2000}. \newline\newline
Because of the maximum distance to Sgr~A* of around 0.1 arcsec, the data selection is limited due to blending events and overlapping orbits of close-by stars (Fig. \ref{fig:all_orbits}). This explains the data point density that is presented in Fig. \ref{fig:orbit}. We conclude, that the observation of S4711 is hindered by the brighter close-by S-stars. Especially close to the pericenter passage of S2 in 2018 \citep{gravity2018b}, a detection of S4711 is not possible.
From the orbit plot of the closest S-cluster members presented in Fig. \ref{fig:all_orbits}, we find an angle between the orbit of S4711 and S62 of about 45$^{\circ}$.
In comparison with the periapsis distance of about 2 mas for S62, we find for S4711 the pericentre distance of
\begin{equation}
    r_{\rm p}\,=\,a(1-e)\,\simeq \,0.7\, mpc
\end{equation}
and using $0.04\,pc\,\approx\,1000\,$mas, we get $r_{\rm p}\,\approx\,12\,mas\,=144\,$AU. Considering the periapsis distance of 120 AU for S2 as obtained by \cite{Gravity2019}, S4711 has a comparable pericentre distance to Sgr~A*. Its apocentre distance is closer than for S2 by $\sim 900\,{\rm AU
}$, see Table~\ref{tab:periapse_shift}. Hence, given the periods of S4711 and S62 shorter than 10 years, these may serve as better probes for relativistic effects than S2.

\subsection{Uncertainties of the orbital parameters}

In \cite{peissker2020a}, we used a variation of the position of S62 with respect to Sgr~A* by $\pm$ 1 px to derive the uncertainties for the orbital elements. However, the orbit of S4711 is even shorter compared to S62. A variation of $\pm$ 1 px overestimates the range of possible orbital solutions compared to the extend of the orbit of S4711. 
\begin{figure*}[htbp!]
	\centering
	\includegraphics[width=1.\textwidth]{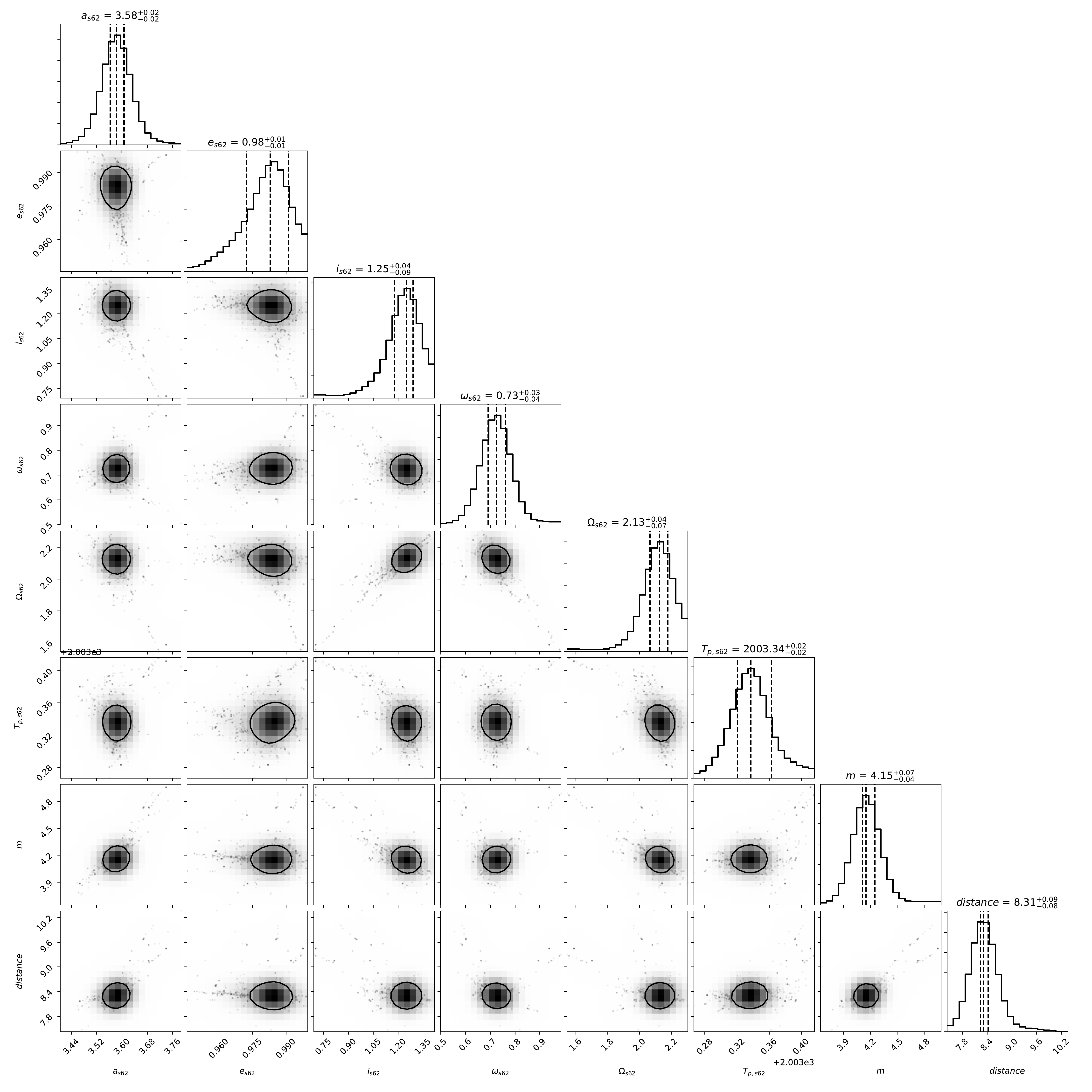}
	\caption{MCMC simulations for the 6 orbital elements based on a Keplerian model for S62. Additionally, two more parameters describing the mass and distance of Sgr~A* are shown. The 1$\sigma$ uncertainty shows throughout the simulations compact shapes. The presented quantiles correspond to the median (0.5) and the 1$\sigma$ central interval (0.16, 0.84).}
\label{fig:mcmc_s62}
\end{figure*}
Hence, we use the Markov chain Monte Carlo (MCMC) affine-invariant ensemble sampler \citep{Foreman-Mackey2013} with PYTHON to derive the uncertainties for S4711, S4712, S4713, S4714, and S4715. For consistency, we also apply the MCMC-Metropolis-Hastings algorithm to the orbital elements of S62 (Fig. \ref{fig:mcmc_s62}). For the apriori distribution (prior), we use the results from the Keplerian fit model (Table \ref{tab:orbital_para}). We then use iteration steps of 100 for the Bayesian updating. With this, we determine the mean of the posterior distribution of the Bayesian inference. For the mass and distance of Sgr A*, we adapt the values $4.15\,\times\,10^{6}M_{\odot}$ and 8.3 kpc from \cite{Parsa2017}, \cite{Gillessen2017}, and \cite{Gravity2019}. 
\begin{figure*}[htbp!]
	\centering
	\includegraphics[width=1.\textwidth]{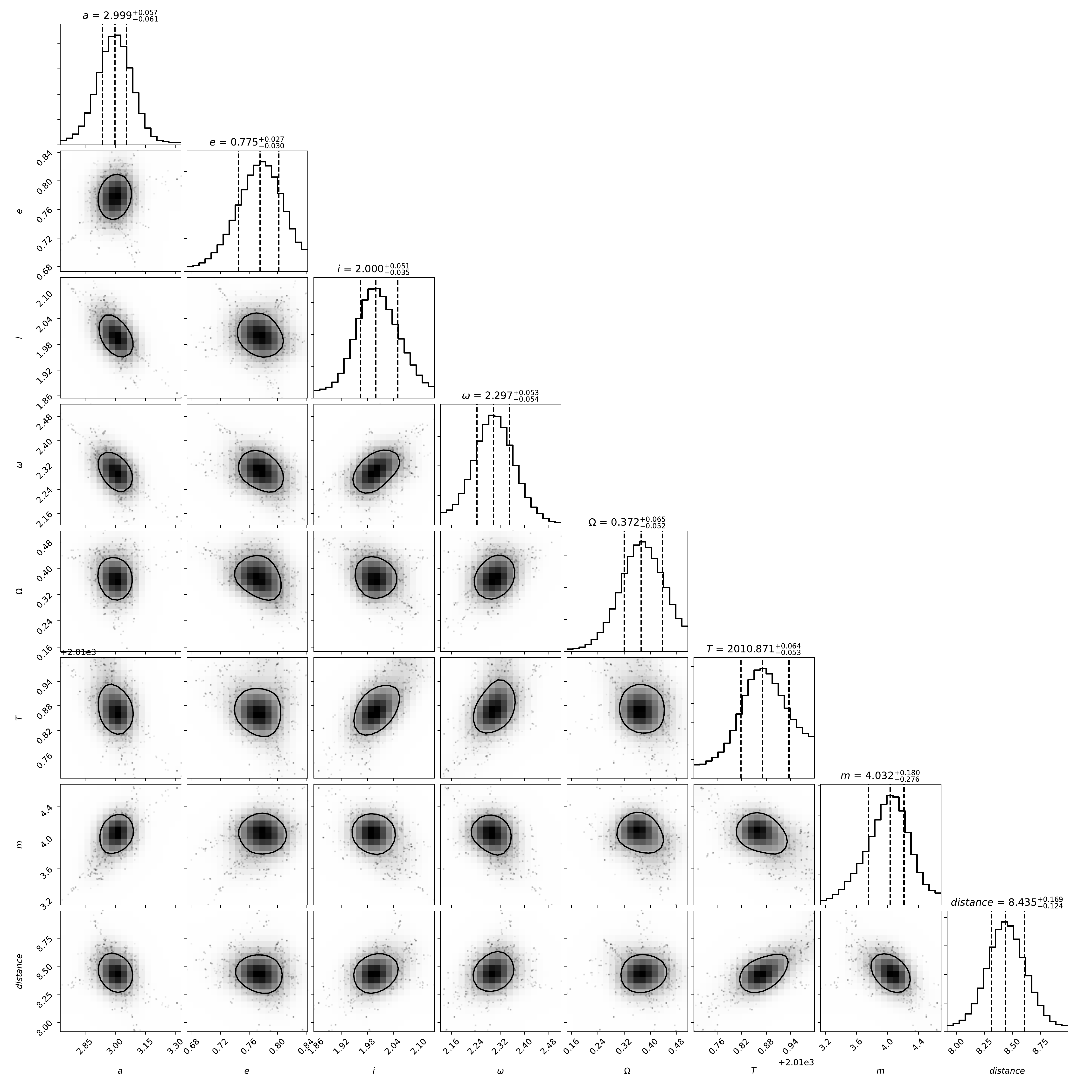}
	\caption{MCMC simulations for the 6 orbital elements based on a Keplerian model for S4711. As mentioned in the text, the mean distribution shows a satisfying agreement with the prior. Additionally to the 6 orbital elements, we also show the MCMC posterior distribution for the mass and distance of Sgr~A*.}
\label{fig:mcmc_s4711}
\end{figure*}
The 1$\sigma$ uncertainties (68$\%$ confidence interval) for the here investigated S-stars are all in a comparable range (see Table \ref{tab:orbital_para}). If the range of the uncertainties for an orbital parameter is not symmetrically distributed, we use the maximum value of the interval. We find, that the MCMC sampler shows a satisfying compact probability for most parameters where the mean is almost identical with the prior. A broader posterior distribution would result in large uncertainties for the orbital parameters.

\subsection{Expected general relativistic effects}
To determine which of the newly identified faint S-stars are good probes of post-Newtonian effects, we first determine the relativistic parameter following \citet{Parsa2017}
\begin{equation}
    \Gamma=\frac{r_{\rm s}}{r_{\rm p}}\,,
\end{equation}
where $r_{\rm s}$ is the Schwarzschild radius of Sgr~A*, $r_{\rm s}=1.21\times 10^{12}(M_{\bullet}/4.1\times 10^6\,M_{\odot})\,{\rm cm}$, and $r_{\rm p}=a(1-e)$ is the pericenter distance of the star. We list the values of $\Gamma$ in Table~\ref{tab:periapse_shift}. We see that S4714 and S62 have the largest values of $\Gamma$, followed by S4711, which has $\Gamma$ comparable to S2. Since the period of S4711 is less than half of the S2 period, its monitoring can reveal post-Newtonian effects twice as fast. The remaining stars, S4712, S4713, and S4715, have $\Gamma$ an order of magnitude smaller than S4711 and hence there are weaker relativistic probes.

Following our analysis in \cite{peissker2020a}, we calculate the Schwarzschild periapsis shift \citep[see][]{Weinberg1972} for S62, S4711-S4715 stars using
\begin{equation}
\delta\phi=\frac{6\pi G}{c^2}\, \frac{M}{a(1-e^2)}=\frac{3\pi}{1+e}\Gamma\,.
\label{eq_periapsis shift}
\end{equation}
In Table~\ref{tab:periapse_shift}, we list the values for all the stars. The largest reliable shift per period is for S62, $\delta \phi = 74.7\pm 31.0$ arcmin \footnote{The value of 9.9$^\circ$ in \cite{peissker2020a} is 
a misprint.}, and for S4714, $\delta \phi = 104.6 \pm 76.3$ arcmin, albeit with a large uncertainty. S4711 has a comparable Schwarzschild precession, $\delta \phi = 10.3 \pm 1.3$ arcmin, to the S-cluster star S2, $\delta \phi = 11.7 \pm 0.6$ arcmin.\newline
\begin{table*}[htbp!]
\centering
        \begin{tabular}{c|c|c|c|c|c|c|c}
        Star  & $r_{\rm p}$ [AU] & $r_{\rm a}$ [AU] & $v_{\rm p}$ [km/s] (\% of c) & $\Gamma [10^{-4}]$ & $z_{\rm gr}c$\,$[{\rm km\,s^{-1}}]$ & $\delta \phi$\,[arcmin] & $\dot{\Omega}_{\rm LT}$\,[arcsec\,${\rm  yr^{-1}}$]\\
        \hline
        \hline
        S62 &  $17.8 \pm 7.4$ & $1462.4 \pm 11.0$ &  $20124 \pm 4244$\, ($6.7 \pm 1.4$) & $46$ & $685.5 \pm 288.6$ & $74.7 \pm 31.0$ & $5.1 \pm 3.2$ \\
        S4711 & $143.7 \pm 18.8$ & $1094.7 \pm 28.7$ & $6693 \pm 494$\, ($2.2 \pm 0.2$) & $5.6$ & $84.5 \pm 11.8$ & $10.3 \pm 1.3$ & $0.34 \pm 0.07$ \\
        S4712 & $2366 \pm 120$   & $5075 \pm 122$  & $1449 \pm 54$\, ($0.48 \pm 0.02$) & $0.34$ & $5.1 \pm 0.4$ & $0.81 \pm 0.05$ & $(5.1 \pm 0.5) \times 10^{-4}$ \\
        S4713 & $1073 \pm 110$ & $2234 \pm 144$ & $2141 \pm 153$\, ($0.71 \pm 0.05$) & $0.75$ & $11.3 \pm 1.3$ & $1.8 \pm 0.1$ & $(5.8 \pm 1.1) \times 10^{-3}$\\
        S4714 & $12.6 \pm 9.3$  & $1670 \pm 10$ & $23928 \pm 8840$\, ($8 \pm 3$) & $64$ & $966.1 \pm 713.5$ & $104.6 \pm 76.3$ & $7.0 \pm 7.6$ \\
        S4715 & $894 \pm 83$ & $1480 \pm 122$ &  $2253 \pm 129$\, ($0.75\pm 0.04$) & $0.90$ & $13.6 \pm 1.4$ & $2.4 \pm 0.2$ & $(1.4 \pm 0.4) \times 10^{-2}$ \\
        \hline 
        S2   & $119.3 \pm 0.3$ & $1949.9 \pm 2.8$ & $7582 \pm 8$\,($2.527 \pm 0.003$) & $6.8$ & $101.7 \pm 5.0$ & $11.7 \pm 0.6$ & $0.19 \pm 0.02$\\
        \hline
        \end{tabular}
    \caption{Additional orbital and relativistic parameters for S62, S4711, S4712, S4713, S4714, S4715 stars in comparison with S2 star. From the left to the right column, we list star name, its pericenter distance (in AU), its apocentre distance (in AU), the pericentre velocity (in km/s and \% of light speed), the relativistic parameter defined as $\Gamma=r_{\rm s}/r_{\rm p}$, the gravitational redshift $z_{\rm gr}c$ (in km/s), the Schwarzschild precession $\delta \phi$ (in arcmin), and the Lense-Thirring precession rate of the ascending node $\dot{\Omega}_{\rm LT}$ for the spin parameter of $0.5$ (in arcsec per yer).}
    \label{tab:periapse_shift}
\end{table*}
Since the first detection of the two stars in 2002 (S62) and 2004 (S4711), several periapse passages have been performed. Because of the periapse shift of S4711, a detectable orbit shift could be observed in the next 10-20 years. In addition, the orbital trajectory of S62 should be shifted by almost 5$^\circ$ between 2023 and 2033. 
Since the bright star S2 is passing by the Sgr~A* region every 15 years, the observation of the periapse passages of S62 and S4711 may be challenging, in particular since one needs to measure orbital sections before and after their passage.

For the remaining faint stars, S4712, S4713, and S4715, we obtained an order of magnitude smaller shifts per period, $\delta \phi=0.81 \pm 0.05$, $1.8 \pm 0.1$, and $2.4 \pm 0.2$ arcmin, respectively, which follows from their values of $\Gamma$ or in other words, from larger pericenter distances due to smaller eccentricities and larger semi-major axes.
%
\begin{figure*}
    \centering
    \includegraphics[width=0.48\textwidth]{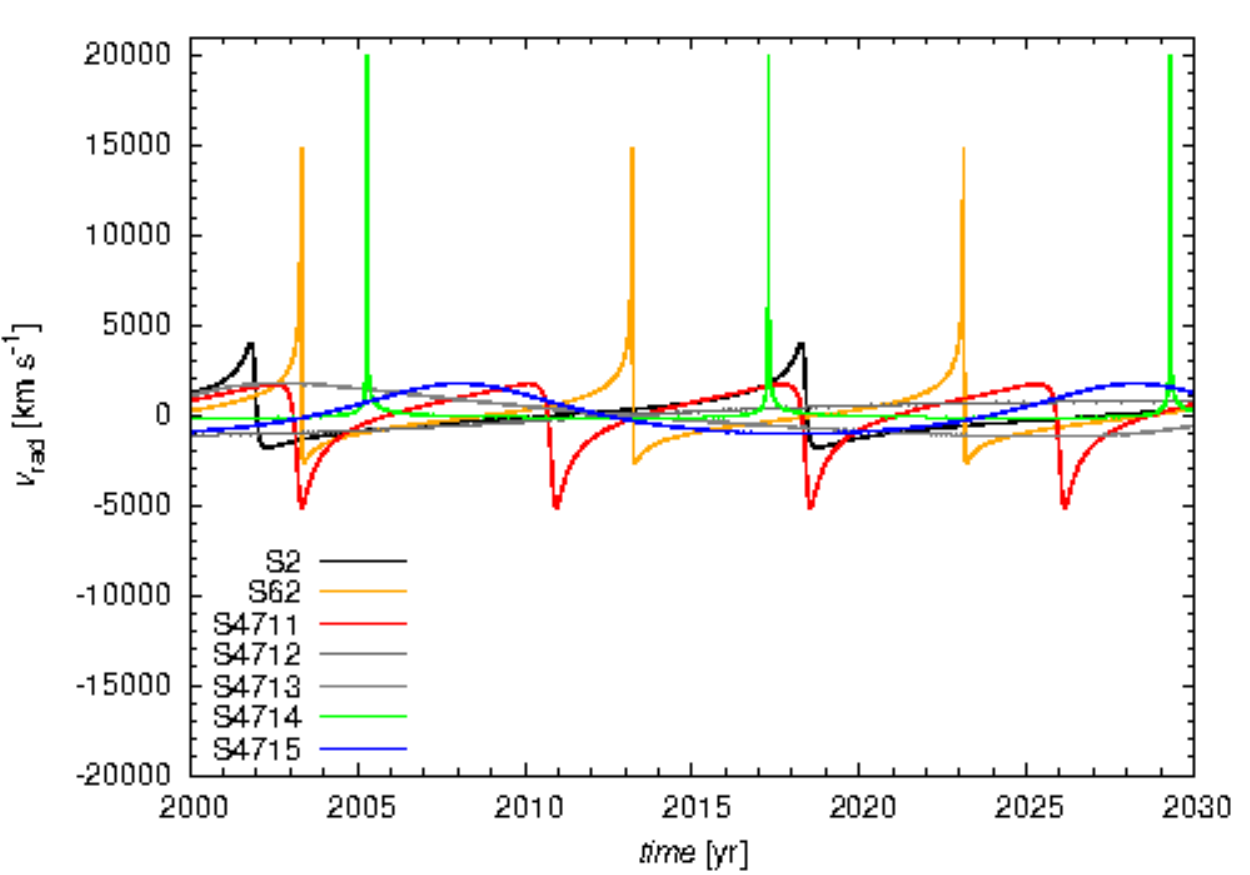}
    \includegraphics[width=0.48\textwidth]{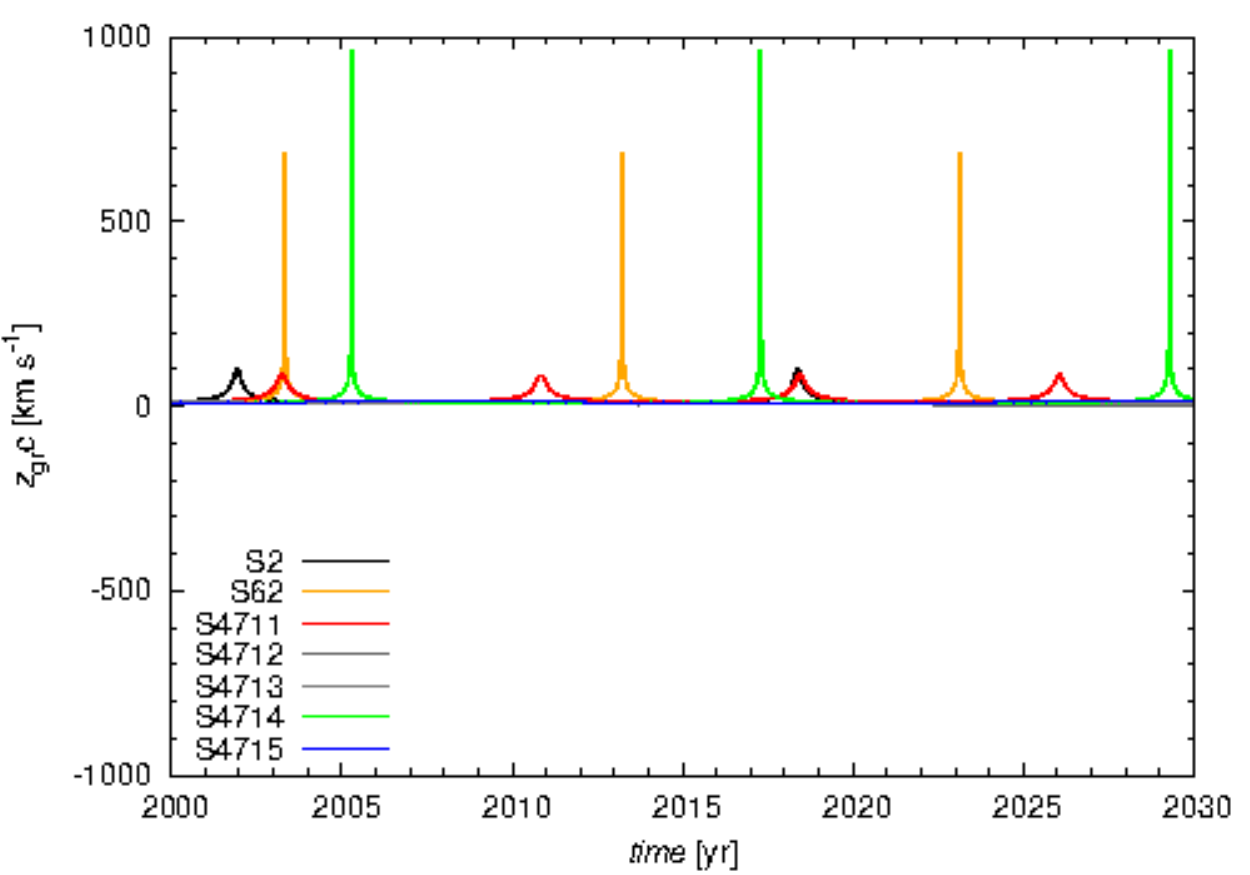}
    \caption{Total radial velocity (left panel) and the gravitational redshift (right panel) expressed in ${\rm km\,s^{-1}}$ as a function of time (in years) calculated based on the orbital elements in Table~\ref{tab:orbital_para} for all the faint stars (S62, S4711, S4712, S4713, S4714, and S4715). S2 is also depicted for comparison.}
    \label{fig_radvel_grshift}
\end{figure*}
Another general relativistic parameter, which affects the observed wavelength of spectral lines, is the gravitational redshift, which can be expressed as
\begin{equation}
   z_{\rm gr}c=c\left[\left(1-\frac{r_{\rm s}}{r} \right)^{-1/2} -1 \right]\lesssim \frac{c\Gamma}{2}\,,
   \label{eq_gr_redshift}
\end{equation}
where the upper limit on the right is the approximate value at the pericentre (the approximation applies for $\Gamma \ll 1$). 
In Fig.~\ref{fig_radvel_grshift}, in the left panel, we show the total radial velocity of the stars, including the radial Doppler shift and the gravitational redshift. In the right panel, we depict the gravitational redshift alone, which is the largest at the pericentre for each star. The values of the gravitational redshift for all the stars are listed in Table~\ref{tab:periapse_shift}. The largest gravitational redshift in the S-cluster so far appears to be for S4714 with $z_{\rm gr}c=966.1 \pm 713.5\,{\rm km\,s^{-1}}$, followed by S62 with $z_{\rm gr}c=685.5 \pm 288.6\,{\rm km\,s^{-1}}$. The gravitational redshift of S4711 $z_{\rm gr}c=84.5 \pm 11.8\,{\rm km\,s^{-1}}$ is within the uncertainty comparable to the value for S2. The remaining faint stars have the gravitational redshift of the order of $10\,{\rm km\,s^{-1}}$.

In addition, for S-stars with pericentre distances of the order of $1000\,r_{\rm s}$ and less, it is of high interest to measure the Lense-Thirring (LT) precession of the ascending node since this would provide an independent probe of the black hole spin \citep{2018MNRAS.476.3600W}. For a star on an elliptical orbit around the SMBH, the LT precession of $\Omega$ can be calculated as follows \citep{2013degn.book.....M}
\begin{equation}
    \dot{\Omega}_{\rm LT}=\frac{2G^2M^2\chi}{c^3a^3(1-e^2)^{3/2}}=\Gamma^2\frac{\chi c}{2r_{\rm a}}\sqrt{\frac{1-e}{1+e}}\,,
    \label{eq_LT_precession}
\end{equation}
where $\chi$ is the dimensionless spin of the black hole ($0<\chi<1$). We list the values of $\dot{\Omega}_{\rm LT}$ in arcseconds per year for all the sources in Table~\ref{tab:periapse_shift} for the assumed spin of $\chi=0.5$, which was chosen based on the spin constraints inferred from modelling the total and polarized flux density of NIR flares \citep[$\chi\geq 0.4$, ][]{2006A&A...460...15M}. The rate of the LT precession is generally larger than for S2 star, in particular we obtained the largest mean value for S4714, $\dot{\Omega}_{\rm LT}=7.0 \pm 7.6\,{\rm arcsec\,yr^{-1}}$, followed by S62 with $5.1 \pm 3.2\,{\rm arcsec\,yr^{-1}}$. S4711 has the larger rate of the LT precession, $\dot{\Omega}_{\rm LT}=0.34 \pm 0.07\,{\rm arcsec\,yr^{-1}}$ than S2 star, which is expected to have $\dot{\Omega}_{\rm LT}=0.19 \pm 0.02\,{\rm arcsec\,yr^{-1}}$. The remaining faint S-stars, namely S4712, S4713, and S4715, have the values of $\dot{\Omega}_{\rm LT}$ at least one order of magnitude smaller.   

However, even for S62 and S4714, which have the best prospects for the Lense-Thirring precession, the monitoring will have to last for at least 20 years to reliably measure the spin. The criterion of \citet{2018MNRAS.476.3600W}, $f_{\rm w}=a(1-e^2)^{3/4}$, is $f_{\rm w}\sim 954\,r_{\rm s}$ for the star S62 and $f_{\rm w}\sim 765\,r_{\rm s}$ for S4714. Following \citet{2018MNRAS.476.3600W}, it would be necessary to monitor S4714 26 years using 120 total observations with $10\,{\rm \mu as}$ astrometric precision to detect the spin of $\chi=0.9$. For S62, the campaign would have to last $\sim 41$ years for the same parameters.

\subsection{Spatial orbital velocity}
According to the Ramanujan approximation of an ellipse \citep[see e.g.,][]{ramanu1988}, the orbital distance U of S4711 travelled during its 7.6 yr orbit can be calculated via
\begin{equation}
    U_{\rm S4711}\,\approx\,(a+b)\,\pi\,\left(1+ \frac{3\lambda^2}{10+\sqrt{4-3\lambda^2}} \right)
\end{equation}
with $\lambda\,=\,\frac{a-b}{a+b}$ where $a$ is taken from Table \ref{tab:orbital_para}. 
The semi-minor axis $b$ can be calculated using the eccentricity $e$ and the semi-major axis $a$ with
\begin{equation}
    b\,=\,\sqrt{a^2-e^2}\,\,\,\,\,\,\, .
\end{equation}
For S4711, we derive a value of $U_{\rm S4711} \approx 18.29$ mpc for the orbital length. In combination with its orbital time period, we derive a mean space velocity for S4711 to around 1540 km/s which equals about 0.5$\%$c. For S62, we derive with the values given in Table \ref{tab:orbital_para} a orbital distance of $U_{\rm S62} \approx 22.11$ mpc with a mean 3d velocity of about 1370 km/s=0.3$\%$c. Applying this analysis to S2 with the orbital elements from \cite{gravity2018} and \cite{Gravity2019}, we get $U_{\rm S2} \approx 31.42$ mpc and a mean 3d velocity of 1915.46 km/s=0.4$\%$c.

\subsection{Indication for a reservoir of faint stars close to Sgr~A*}

\begin{figure}[htbp!]
	\centering
	\includegraphics[width=.5\textwidth]{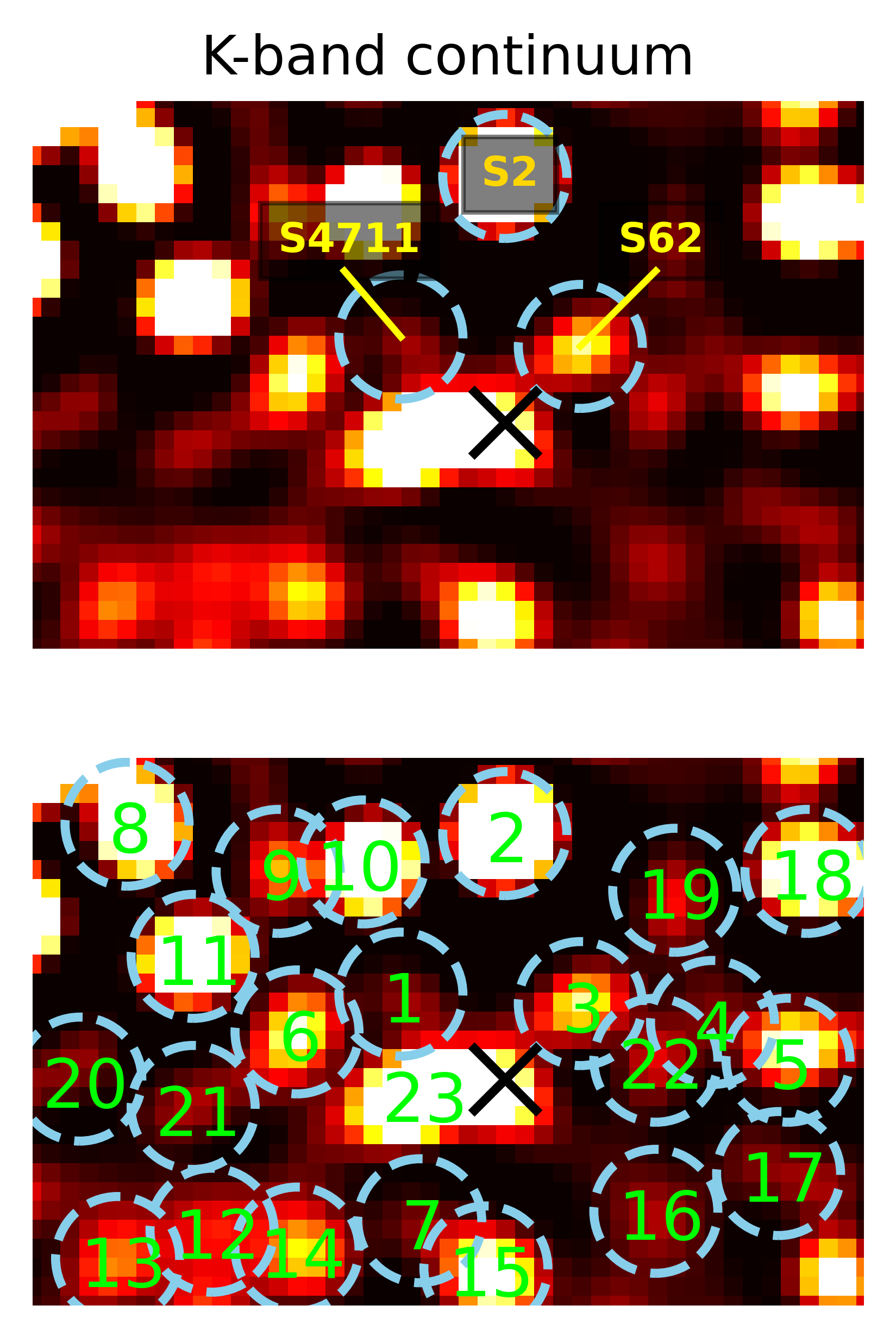}
	\caption{Finding chart of the 25 innermost objects around Sgr~A* based on the NACO GC data of 2008. Both images show the same data. We mark S2, S62, and S4711 in the upper image. In the lower one, we mark all visible objects with an id. The related objects are in consecutive order: 1/S4711, 2/S2, 3/S62, 4/S4712, 5/S38, 6/S40, 7/S4713, 8/S14, 9/S56, 10/S13, 11/S20, 12/DSO, 13/S23, 14/S54, 15/S19, 16/S4718, 17/S61, 18/S31, 19/S4714, 20/a, 21/b, 22/S4715. Next to the black x (Sgr~A*), the id $\#$23 marks the position of S17, S55, and S175. The objects 20/a and 21/b can be observed at the same position relative to Sgr~A* between 2006 and 2015 and are therefore foreground or background sources.}
\label{fig:naco_2008}
\end{figure}

The analysis of the NACO and SINFONI data reveals not only the existence of S4711, but also shows that there are several FSS-candidates (Fig. \ref{fig:naco_2008}).
The results of the observation of these FSS candidates are summarized in the following. For clarity, we follow the here introduced nomenclature and call them S4712, S4713, S4714, and S4715 to underline their uniqueness.

{\it S4712:}
This S-cluster member can be observed between 2008 and 2015. Because of S2, S19, and S13, a detection before 2008 and after 2015 is not free of confusion. The orbital elements are based on the Keplerian fit of the NACO data and are listed in Table \ref{tab:orbital_para}.

{\it S4713:}
Compared to S4712, the observation of S4713 is limited to the years 2006 to 2010. Like for S4712, the stars S2, S19, S55, and S175 hinders a confusion free observation of S4713 before and after 2006 and 2010.

{\it S4714:}
The faint star S4714 can be observed for 4 consecutive years. In contrast to S4712 and S4713, the orbital time period is much smaller and is determined to be around 12 years (see Table \ref{tab:orbital_para}). The magnitude of the orbital time period is in line with S55 and S62. These two S-stars are also hindering the observation of S4714 because of blending effects.

{\it S4715:}
S4715 has a longer orbital period than S2 star, which is $20.2 \pm 2.4$ years. Together with S4712 and S4713, these are longer-period newly discovered faint S stars in comparison with S62, S4711, and S4714. Because of the crowded S-cluster, it should be noted that we do not observe a full orbit of S4712, S4713, S4714, and S4715. Because of the observation in several consecutive years, a blend star scenario \citep{Sabha2012, peissker2020a} can be excluded.

{\it S4716:}
As shown in Fig. \ref{fig:naco_2008}, we find another S-cluster member between S19 and S38 that we name S4716 to underline the character of the source. However, we observe this object in the NACO data set of 2008 and 2009. Because of blending and the overlap of other nearby stars, an observation before 2008 and after 2009 is not satisfying.

As for S4712, S4713, S4714, and S4715, it is unlikely that these stars are blended background or foreground stars. This is underlined by the observation of more than 3 consecutive years \citep{Sabha2012, peissker2020a}. When it comes to 20/a, 21/b, and S4716, it could be argued that these stars are rather fore/background stars that are passing Sgr~A* at small projected distances \citep[see][]{Ghez2002, Eckart2002}.

\subsection{Eccentricity relation}
\begin{figure}[htbp!]
	\centering
	\includegraphics[width=.5\textwidth]{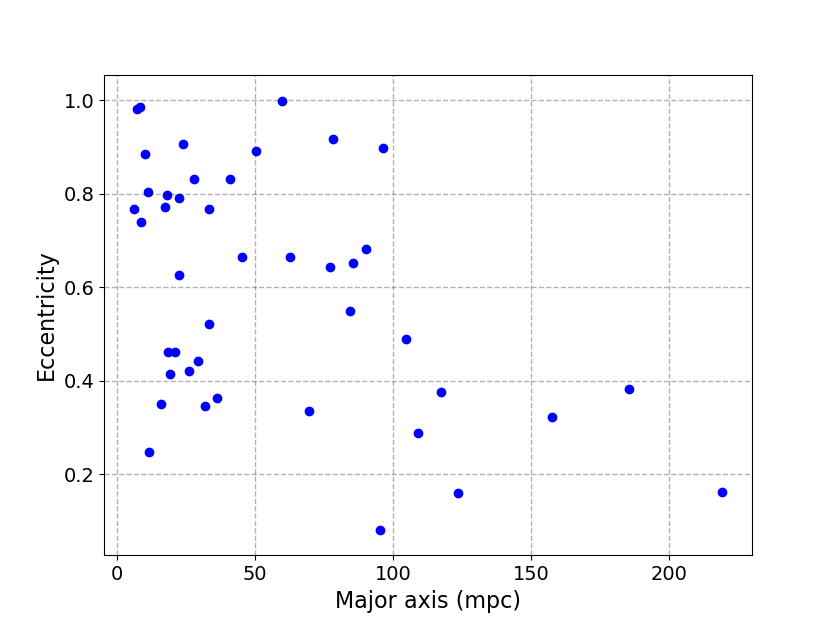}
	\caption{Eccentricity of all known S-stars as a function of the major axis. The used values for the plot are based on a Keplerian fit. The related orbital elements are presented in  \cite{Ali2020}.}
\label{fig:eccentricity_semi_major}
\end{figure}

\cite{peissker2020a}, \cite{Ali2020}, and this work investigate members of the S-cluster. S62, S4711, and S4714 show highly eccentric orbits with a short semi-major axis a (see Table \ref{tab:orbital_para}). In Fig. \ref{fig:eccentricity_semi_major}, we show the relation between the eccentricity as a function of the major axis.
We want to point out a not shown outlier in Fig. \ref{fig:eccentricity_semi_major}. This outlier is S85 and is 144 mpc away from the remaining stars. This is a significant distance relative to the extent of the S-cluster of around 40 mpc. Considering the large distance from the inner arcsecond, it raises doubt about the classification of S85 as a member of the S-cluster. In the light of the Hills mechanism \citep{Hills1988} and the work of \cite{Zajacek2014} that are mentioned in the introduction, a detailed analysis of S85 could be fruitful.
Besides the outlier, the trend of high eccentricity followed by low major-axis is detected for the S-stars. We find for all stars with a larger major axis ($>$ 100 mpc) eccentricity values below 0.5. Vise versa, the density distribution of S-stars with a major axis $<$ 100 mpc shows a tendency towards eccentricity values higher than 0.5.
\begin{figure}[htbp!]
	\centering
	\includegraphics[width=.5\textwidth]{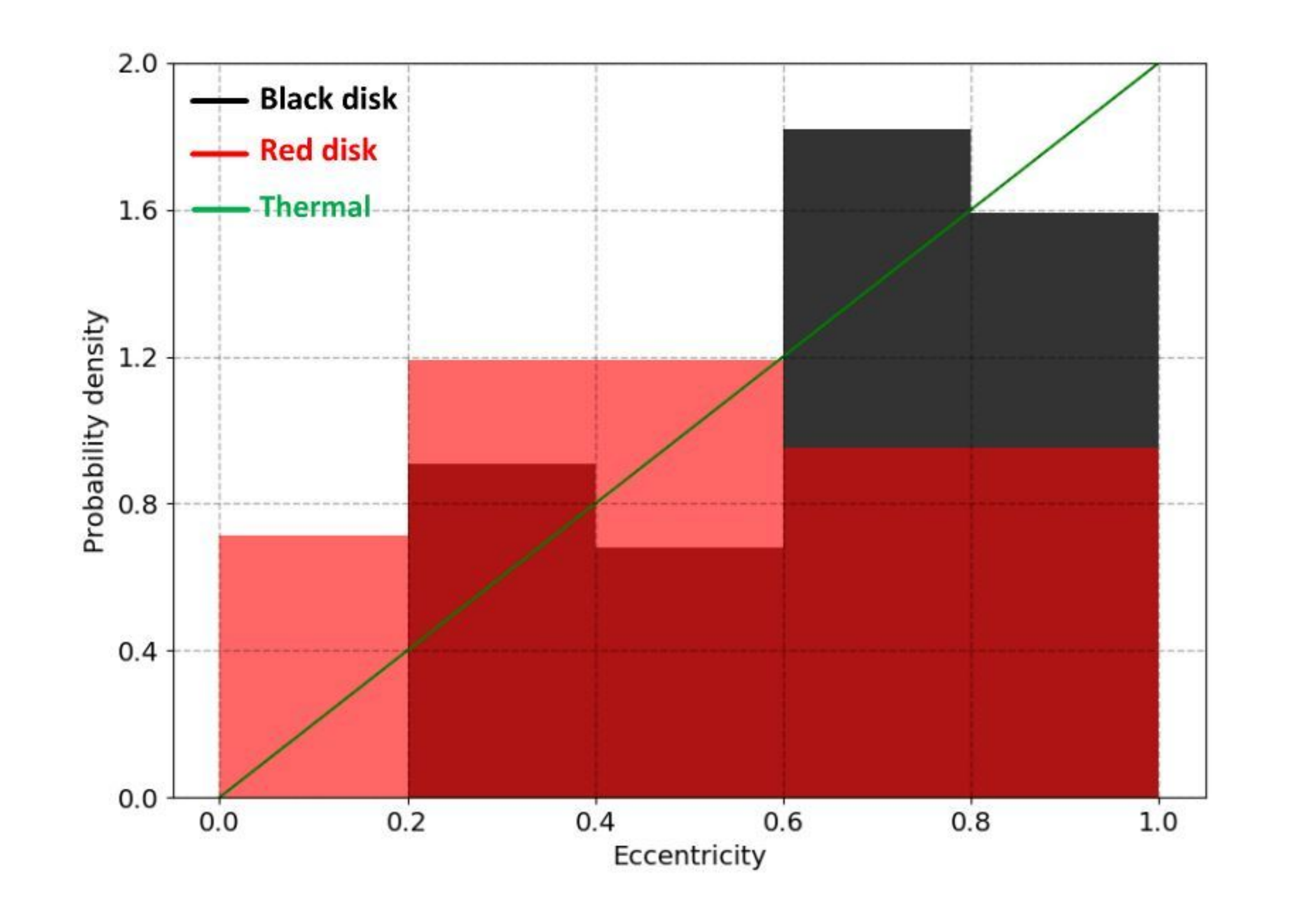}
	\caption{Distribution of eccentricities of both stellar disks identified by \cite{Ali2020} including the newly identified stars. Dark red indicates the overlap between the red and the black disk in this diagram.
}
\label{fig:eccenall}
\end{figure}
\newline 
By examining Fig. \ref{fig:eccenall} that shows the eccentricity distribution of known S-stars orbits, including the newly identified sources, we find that the more compact black disk \citep[see][]{Ali2020} exhibits a thermalized distribution \citep[$f(e) \times de = 2e \times de$,][]{Jeans}. Here, the term "thermalized" refers to reaching statistical equilibrium as a result of stellar interactions. The reason behind this may be the result of many Kozai-Lidov cycles that are triggered with the presence of a massive disturber \citep[see][]{Chen, Ali2020}. Furthermore, it can't be excluded that the Hills mechanism could also be a contributing process, especially for $e > 0.9$. It's worth noting that the short orbital timescale of these inner stars is probably the reason why they reached the thermalized arrangement in a short time. 
In contrast, the red disk displays rather a nearly flat distribution, peaking at $e = 0.4$, i.e., not thermalized. This could be due to the longer orbital periods and the on-average larger distances from Sgr~A*. Therefore, the red disk stars might still need longer time to achieve the thermalized case. Nevertheless, they could also be subject to disk-migration process that leaves them in a non-thermalized state.
\newline 
In addition to the previous possibilities, \cite{Merritt} proposes a plausible explanation for the current distribution of the S-stars. Loss-cone dynamics describe the process in which the star suffers loss of angular momentum due to gravitational radiation at a very close distance to the SMBH. This leads to an orbit with high eccentricity. It is possible in this scenario that the star ends up captured by the central mass. On the contrary, the star could evolve after some time and cross the Schwarzschild barrier, which defines the line of maximum eccentricity in the $e-a$ diagram. As soon as this is achieved, resonant relaxation (RR) is then triggered, allowing the star to gain angular momentum and thus lowering its orbital eccentricity. 
\newline 
Based on this argument, we conclude that the red disk may be relaxed by RR, while the black disk is subject to angular momentum loss process. However, in \cite{Ali2020} they find that the timescale of the scalar resonant relaxation that changes the value of angular momentum exceeds the current estimated age of the S-stars. Regarding this fact and using Newtonian simulations, \citep{Perets} show  that RR could already exhibits its dynamical effects in case the nucleus has a large density of compact disturbers (e.g. a population of black holes). Therefore, along with the previously mentioned possibilities, the proposed loss-cone justification could be considered as a valid scenario that describes the current eccentricity evolution.

\subsection{Kinematic structure of the central S-cluster}

\cite{Ali2020} present a detailed analysis of the kinematics of 112 stars
that are present in the high velocity S-cluster and orbit the super massive black hole Sgr~A*.
They find, that the distribution of inclinations and flight directions
deviate significantly from a uniform distribution that one would expect in the case of random orientation of the orbits.
The S-cluster stars are arranged in two almost edge on thick disks that are located at a position angle about $\pm$45$^o$ with respect to the Galactic plane.
With 25$^o$, the poles of this structure are close to the line of sight. The reason for this arrangement is unclear.
It may be due to a resonance process that started during the formation
of the cluster of young B-dwarf stars about 6~Myr ago. Another possibility is
the presence of a disturber at a distance of a few arcseconds from the S-cluster.

In relation to these two disks identified by \cite{Ali2020},
we find that the stars S4711, S4712, and S4713 belong to the more compact
(about $<$0.5'' diameter) black disk, while S4714 and S4715 are members of the
larger ($\sim$1'' to 2'' diameter) red disk.
More precisely and according to the three-dimensional visual
inspection, the orbit of S4715 seems to have approximately
a 30$^o$ deviation from the plane of the red disk.
As for the remaining newly identified stars, their orbits
fit on average to within about 10$^o$ into the two disk scheme.
This implies that even the stars currently closest (both, physically and
in projection) to SgrA* follow the kinematic structure of the S-cluster.

\section{Discussion \& Conclusion}
\label{sec:discussion}
In this work, we derive various properties for the newly discovered faint 
star S4711 that we detected with SINFONI and NACO in several consecutive years.
S4711 is on a 7.6 year orbit around the SMBH Sgr~A* with a high eccentricity of $e\,=\,0.768$. With a K-band magnitude of about 18.3 it is among the fainter S-cluster members.
The mean spatial orbital velocity of S4711 of about 0.5$\%$c and is in line with other 
S-stars members like S62 and S2 close to Sgr~A*. 
We can conclude, that S4711 is without a doubt a member of the S-cluster. This is followed by the detection of S4712-S4715. From these stars, S4714 is the most interesting and shows comparable properties to S4711. The S-cluster star S4714 shows an even higher eccentricity than S62 of $e\,=\,0.985$. With this, S4714 is the star with one of the highest eccentricity of the S-cluster except for S175.
Given the detection of S62, S4711, and S4714 we find a population of stars 
on orbits smaller than that of S2. As a result of improved data analysis,
we expect to find more stars of this class in the near future. The observation of S62, S4711, and S4714 as well as the increased probability of finding even closer stars on shorter orbits in the future with the Extremely-Large-Telescope ($\sim\,40\,m$ diameter) underlines the prediction of \cite{Alexander2003a}, where the authors propose the existence of the so-called squeezars. This class of stars are on highly eccentric orbits around a super massive black hole and can be categorized as hot and cold squeezars (HS and CS respectively). Additionally, the authors of \cite{Alexander2003b} describe orbital in-spiral processes. Together with \cite{Alexander2003a}, the high eccentricity can be an indication of a tidal disruption event resulting in an increased flare activity. The statistical behavior of Sgr~A* is investigated by \cite{Witzel2012} and with this model, upcoming observations should trace footprints of a tidal disruption event. We speculate, that the unusual strong IR flare observed by \cite{Do2019} could be linked to such an event.

However, additional monitoring data will help to discriminate the faint stars S4711-S4715 from the brighter cluster members in the nuclear region in which the arrangement
of cluster members is changing on a weekly to yearly time scale. This could also help to confirm the relativistic periapse shift determined for S62 and S4711. Only a large coverage of the orbit can help to distinguish uncertainties from relativistic orbital parameter. Since we already observed a full orbit for S62 between 2003 and 2013, we should be able to see a relativistic footprint in the orbit of this S-star.


Furthermore, by analysing the positional data of 20/a, 21/b (see Fig. \ref{fig:naco_2008}), and S4716 we do not find a notable curvature nor a significant acceleration. This is suspicious considering the short projected separation of these FSS from Sgr~A*. It is more likely, that at least 20/a and 21/b are not members of the central part of the S-cluster but rather belong the 
the overall Galactic center stellar cluster. Because of the limited detection of S4716 for about 2 years we speculate a blend star event even though we detect this object in individual data-sets throughout 2008 and 2009. 

As discussed before, we expect several new findings of close stars/squezzars with NIR observations carried out with the ELT in the future. This assumption is based on the observation of S62, S4711, and S4714. The existence of squezzars in combination with unusual high flaring activities of Sgr~A* could be an interesting part of a scientific framework.

\acknowledgments

We would like to thank the anonymous referee for the encouraging and positive comments. With the help of these comments, we were able to improve the paper.
We thank the members of the NACO/SINFONI and ESO Paranal/Chile team. This work was supported in part by the Deutsche Forschungsgemeinschaft (DFG) via the Cologne Bonn Graduate School (BCGS), the Max Planck Society through the International Max Planck Research School (IMPRS) for Astronomy and Astrophysics, as well as special funds through the University of Cologne and SFB 956: Conditions and Impact of Star Formation. We thank the Collaborative Research Centre 956, sub-project [A02], funded by the Deutsche Forschungsgemeinschaft (DFG) project ID 184018867. MZ acknowledges the financial support by the National Science Centre, Poland, grant No.~2017/26/A/ST9/00756 (Maestro 9).
\facilities{VLT(SINFONI), VLT(NACO), VLT(GRAVITY)}

\newpage 
\appendix

\section{Stellar motion of S4712, S4713, S4714, and S4715}
\begin{figure}[htbp!]
    \centering
    \begin{minipage}{0.45\textwidth}
        \centering
        \includegraphics[width=0.9\textwidth]{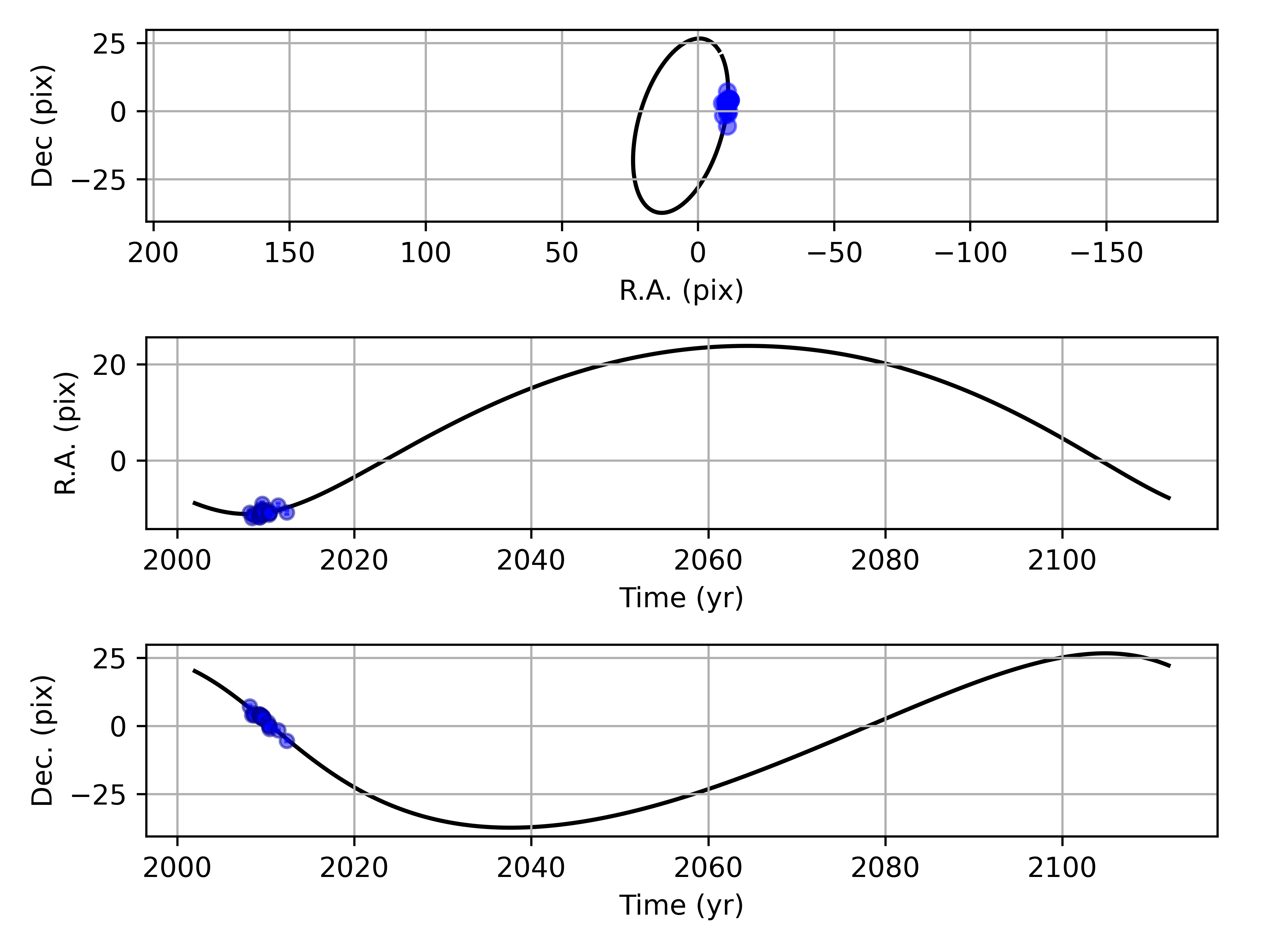} 
        \caption{Detection of the stellar orbit of S4712 based on NACO data of 2008-2015. The size of the data points are larger than the typical spatial uncertainties of $\pm$0.5 px. One full orbit is presented in this figure.}
        \label{fig:orbit_S4712}
    \end{minipage}\hfill
    \begin{minipage}{0.45\textwidth}
        \centering
        \includegraphics[width=0.9\textwidth]{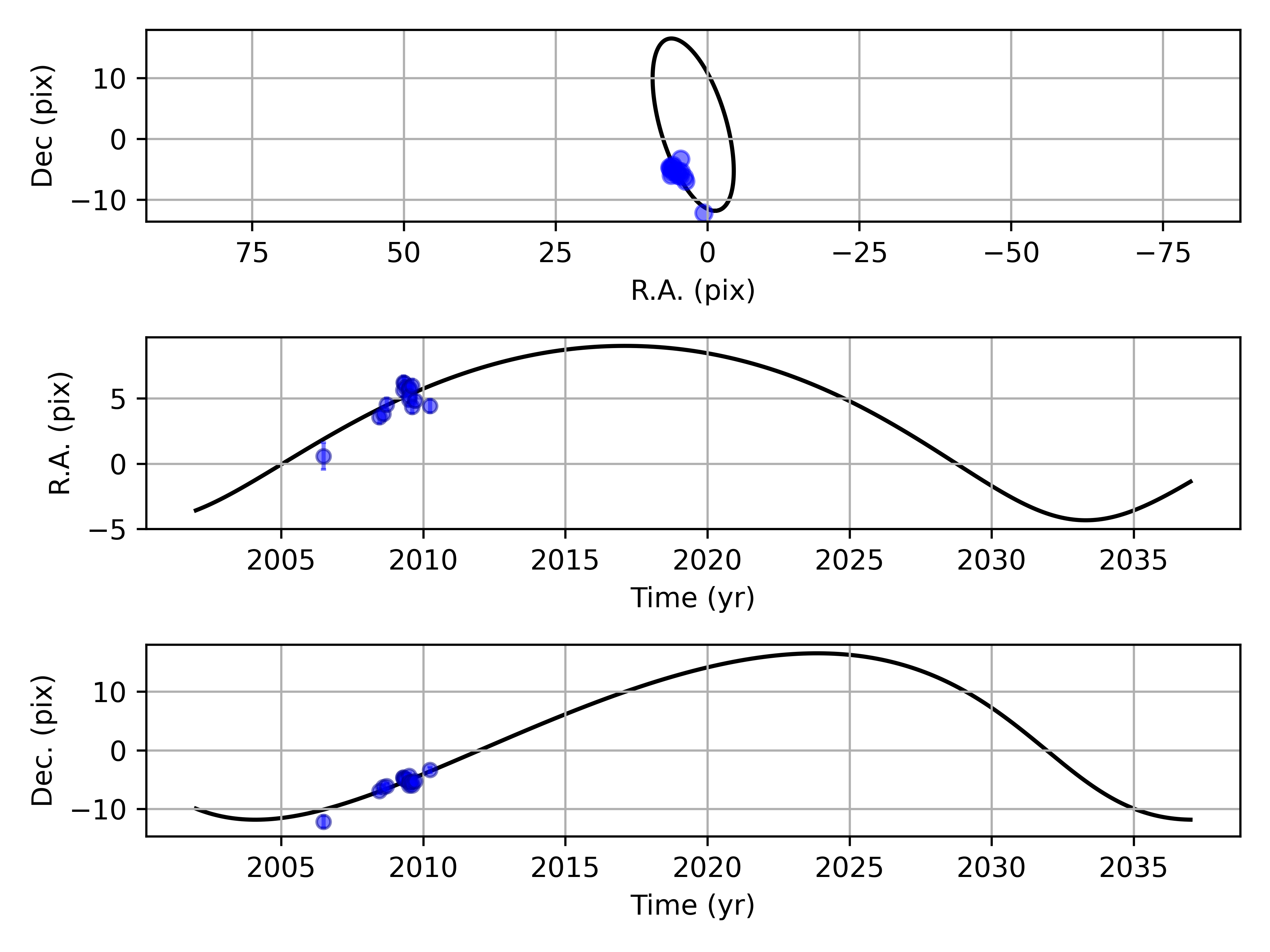} 
        \caption{S4713 detected with NACO between 2006 and 2010. Except for the R.A. data point in 2006, the typical uncertainties are $\pm$0.5 px. In 2006, the presence of the close by S-stars S17, S19, and S55 results in an increased spatial uncertainty of $\pm$1 px.}
        \label{fig:orbit_S4713}
    \end{minipage}
\end{figure}

\begin{figure}[htbp!]
    \centering
    \begin{minipage}{0.45\textwidth}
        \centering
        \includegraphics[width=0.9\textwidth]{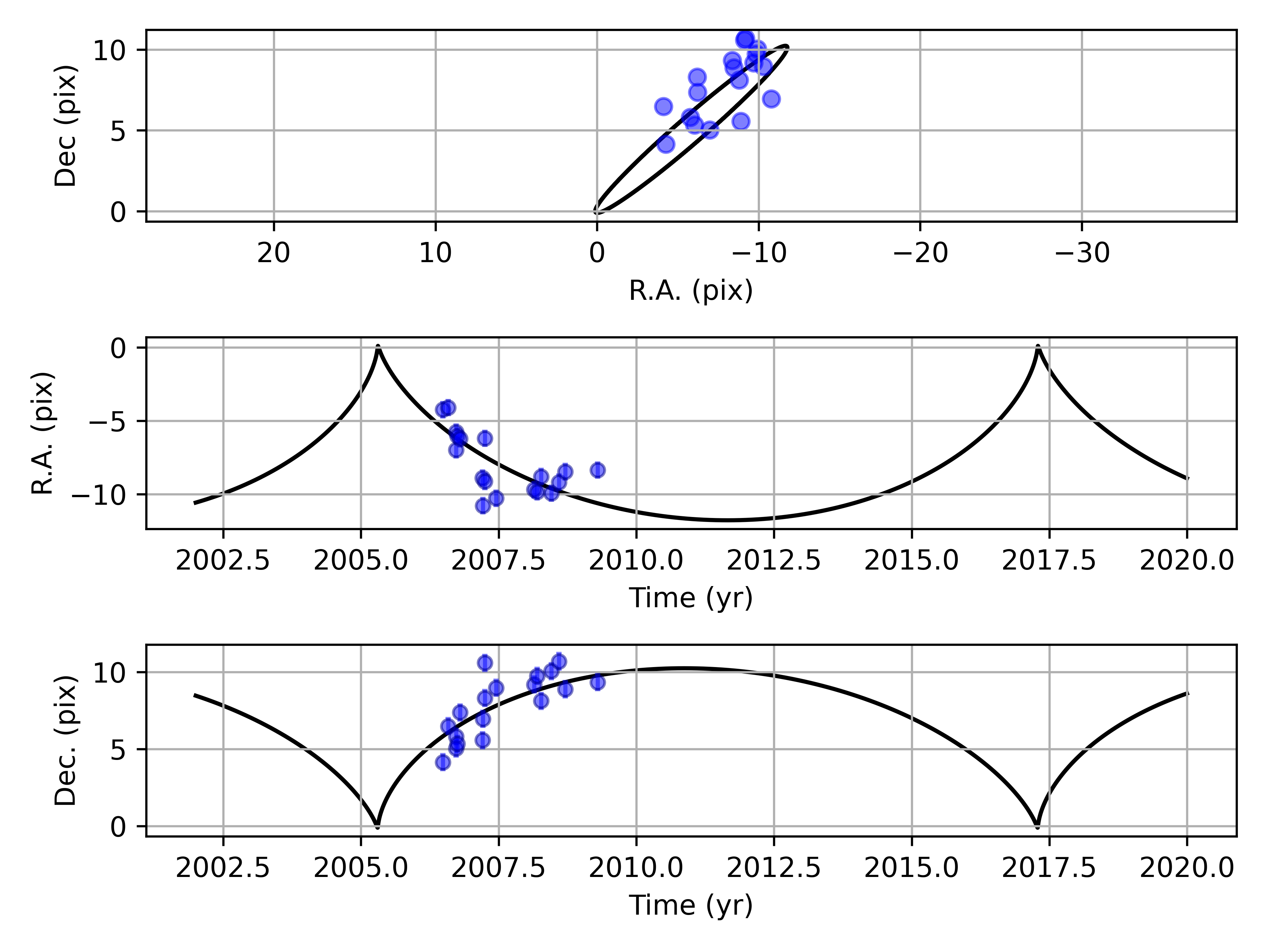} 
        \caption{Detection of the stellar orbit of S4714 based on NACO data of 2006-2009. Like S55, S62, and S4711, the orbital period of S4714 is shorter compared to S2.}
        \label{fig:orbit_S4714}
    \end{minipage}\hfill
    \begin{minipage}{0.45\textwidth}
        \centering
        \includegraphics[width=0.9\textwidth]{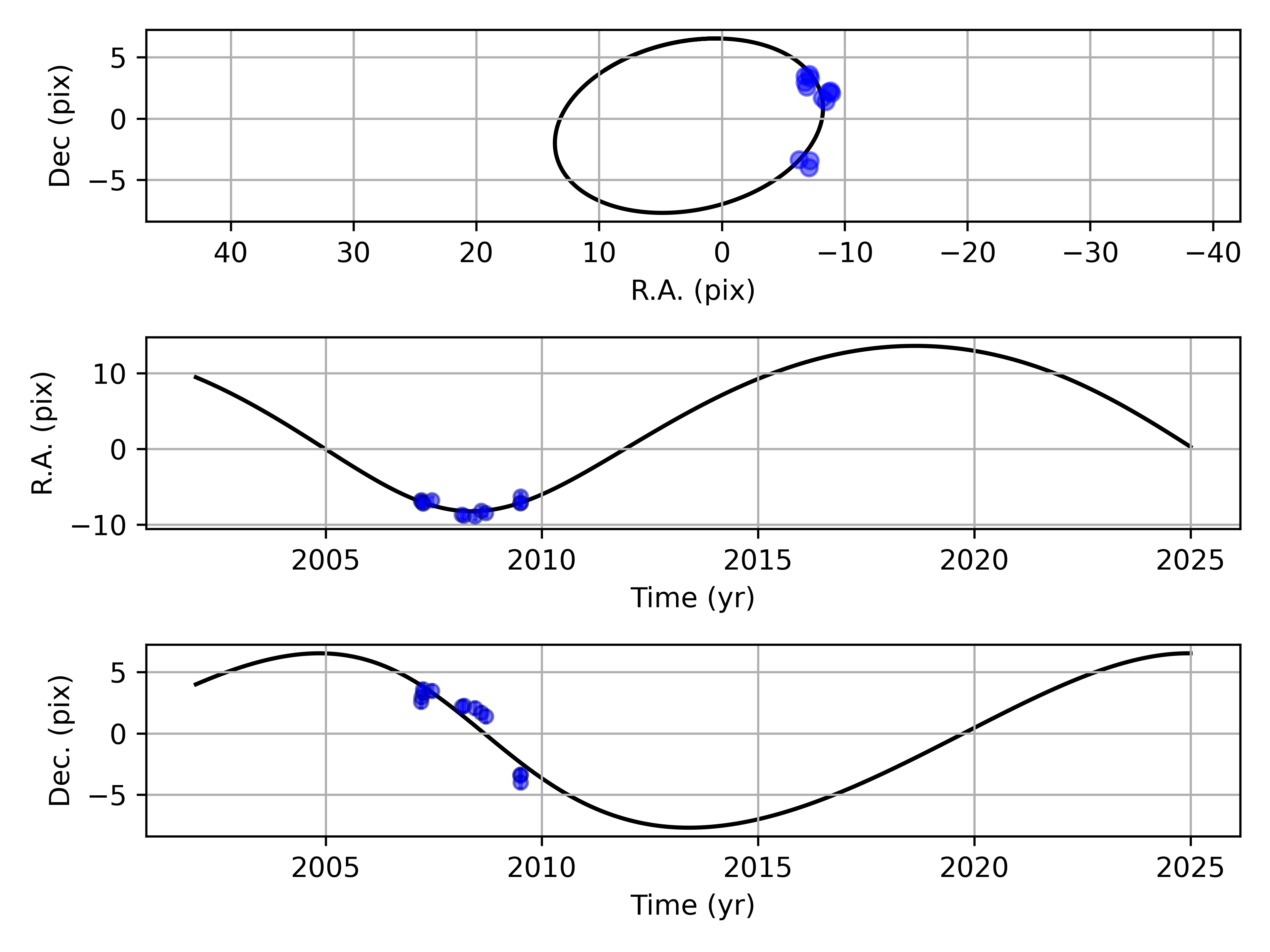} 
        \caption{S4715 detected with NACO between 2007 and 2009. The uncertainties are as big as the data points and in the typical range of $\pm\,0.5$ px or $\pm\,6.5$ mas.}
        \label{fig:orbit_S4715}
    \end{minipage}
\end{figure}

\begin{figure*}[htbp!]
	\centering
	\includegraphics[width=1.\textwidth]{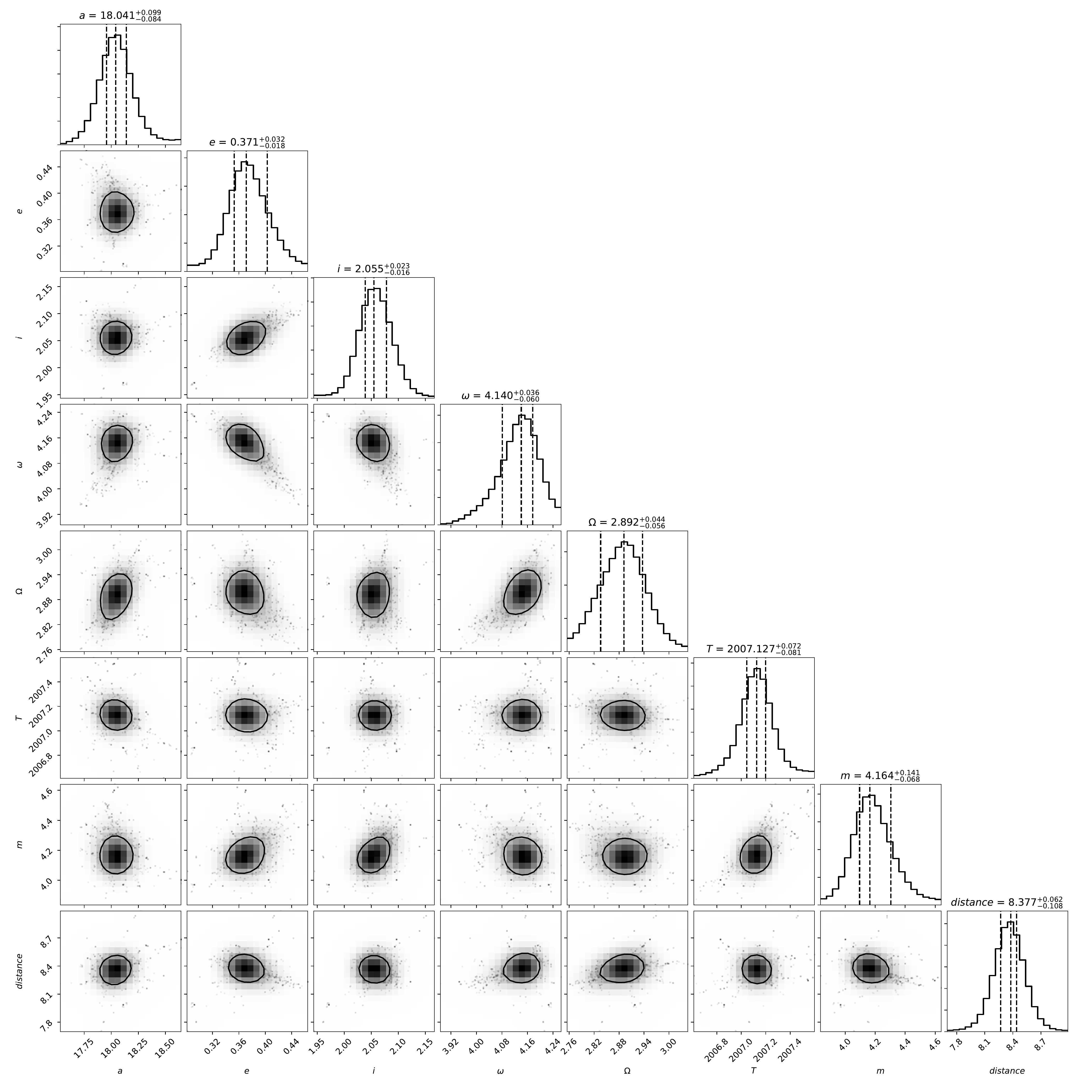}
	\caption{MCMC simulations of S4712. Like before, we additionally show the posterior distribution for the mass and distance of Sgr~A*. The compact distribution of the shown parameter indicates a satisfying orbital analysis.}
\label{fig:mcmc_s4712}
\end{figure*}

\begin{figure*}[htbp!]
	\centering
	\includegraphics[width=1.\textwidth]{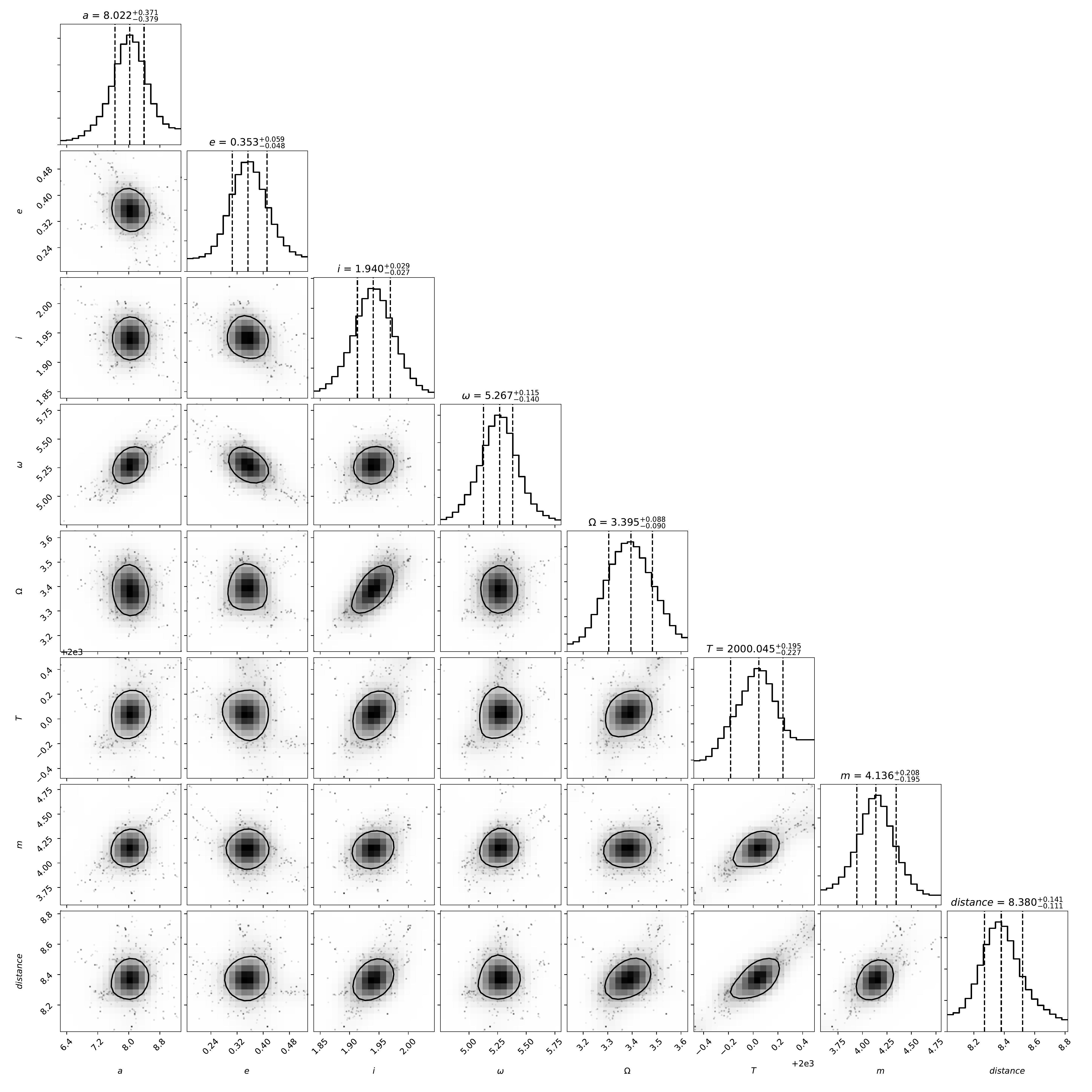}
	\caption{MCMC simulations of S4713.}
\label{fig:mcmc_s4713}
\end{figure*}

\begin{figure*}[htbp!]
	\centering
	\includegraphics[width=1.\textwidth]{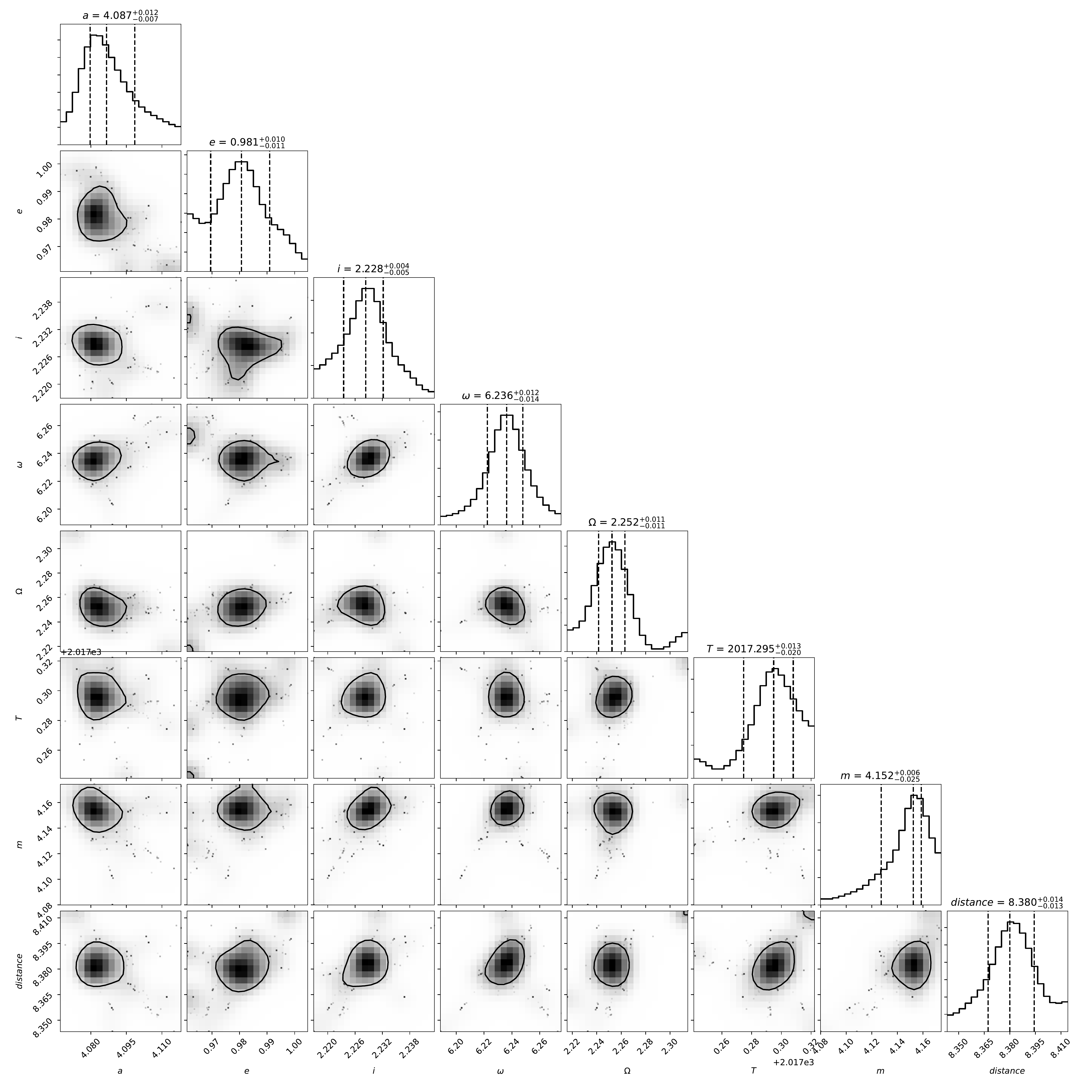}
	\caption{MCMC simulations of S4714.}
\label{fig:mcmc_s4714}
\end{figure*}

\begin{figure*}[htbp!]
	\centering
	\includegraphics[width=1.\textwidth]{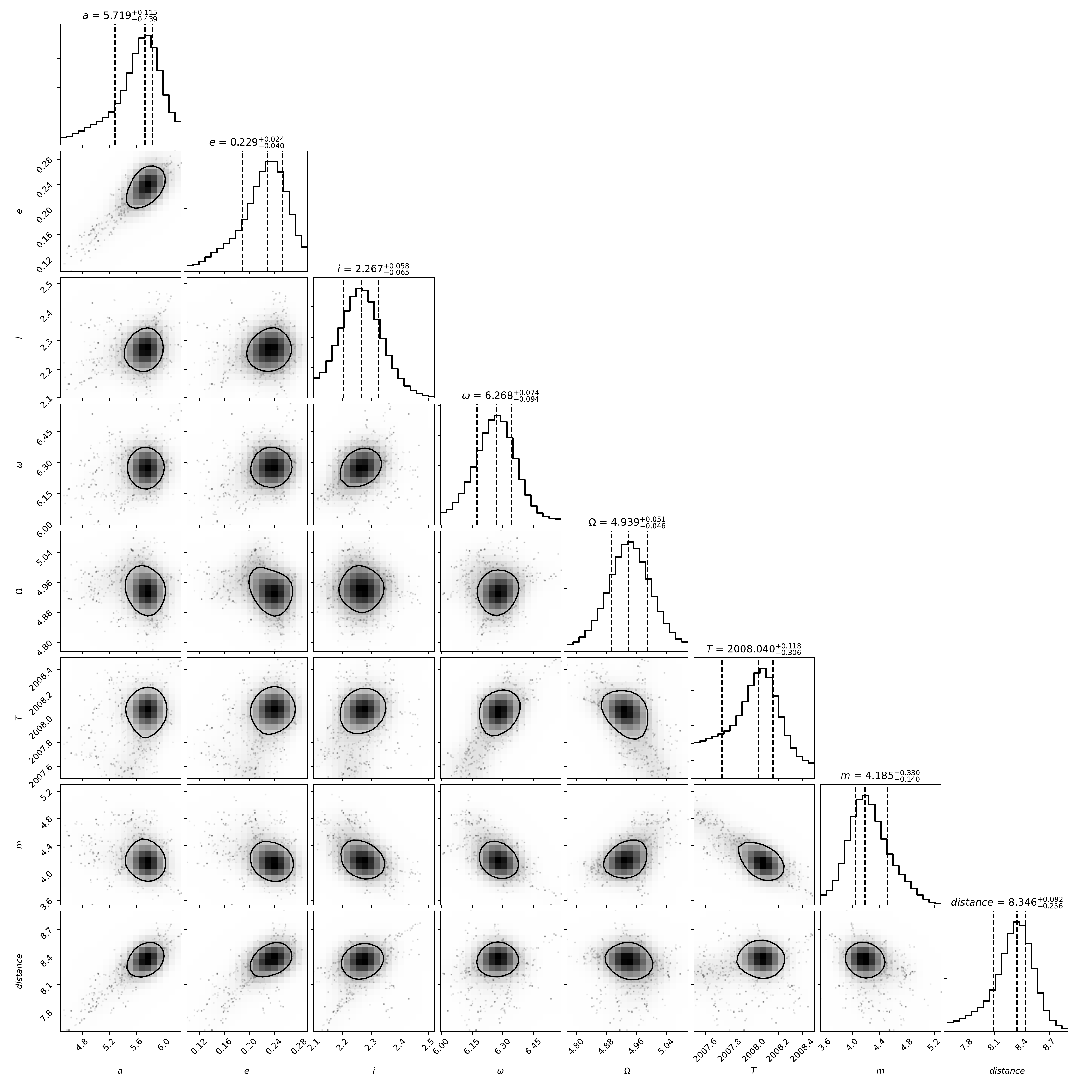}
	\caption{MCMC simulations of S4715.}
\label{fig:mcmc_s4715}
\end{figure*}

In this section, we present the orbits of S4712 (Fig. \ref{fig:orbit_S4712}), S4713 (Fig. \ref{fig:orbit_S4713}), S4714 (Fig. \ref{fig:orbit_S4714}), and S4715 (Fig. \ref{fig:orbit_S4715}). For the data-points, typical uncertainties of $\pm\,6.5$ mas are used. They are fully compatible with the analysis of S62 \citep{peissker2020a} where we use these uncertainties for the data representation in the corresponding figures. The coverage of a full orbit is not given (see Sec. \ref{sec:results}). The orbital elements with the related uncertainties that are based on the MCMC simulations (Fig. \ref{fig:mcmc_s4712}, \ref{fig:mcmc_s4713}, \ref{fig:mcmc_s4714}, \ref{fig:mcmc_s4715}) can be found in Table \ref{tab:orbital_para}. For S4714, we find an orbit that is comparable to S55 when it comes to the orbital period.  

Because of the crowded FOV, observing S4712, S4713, and S4715 is challenging. However, we trace these three S-cluster members in the years between 2008-2015 (S4712), 2006-2010 (S4713), and 2007-2009 (S4715). Since bright S-stars like S2, S13, S19, or S31 are contaminating the orbits of S4712, S4713, and S4715, an observation before 2006 and after 2015 is not free of confusion.

\section{BONNSAI simulations of the S-star S4711}
\begin{figure*}[htbp!]
	\centering
	\includegraphics[width=1.\textwidth]{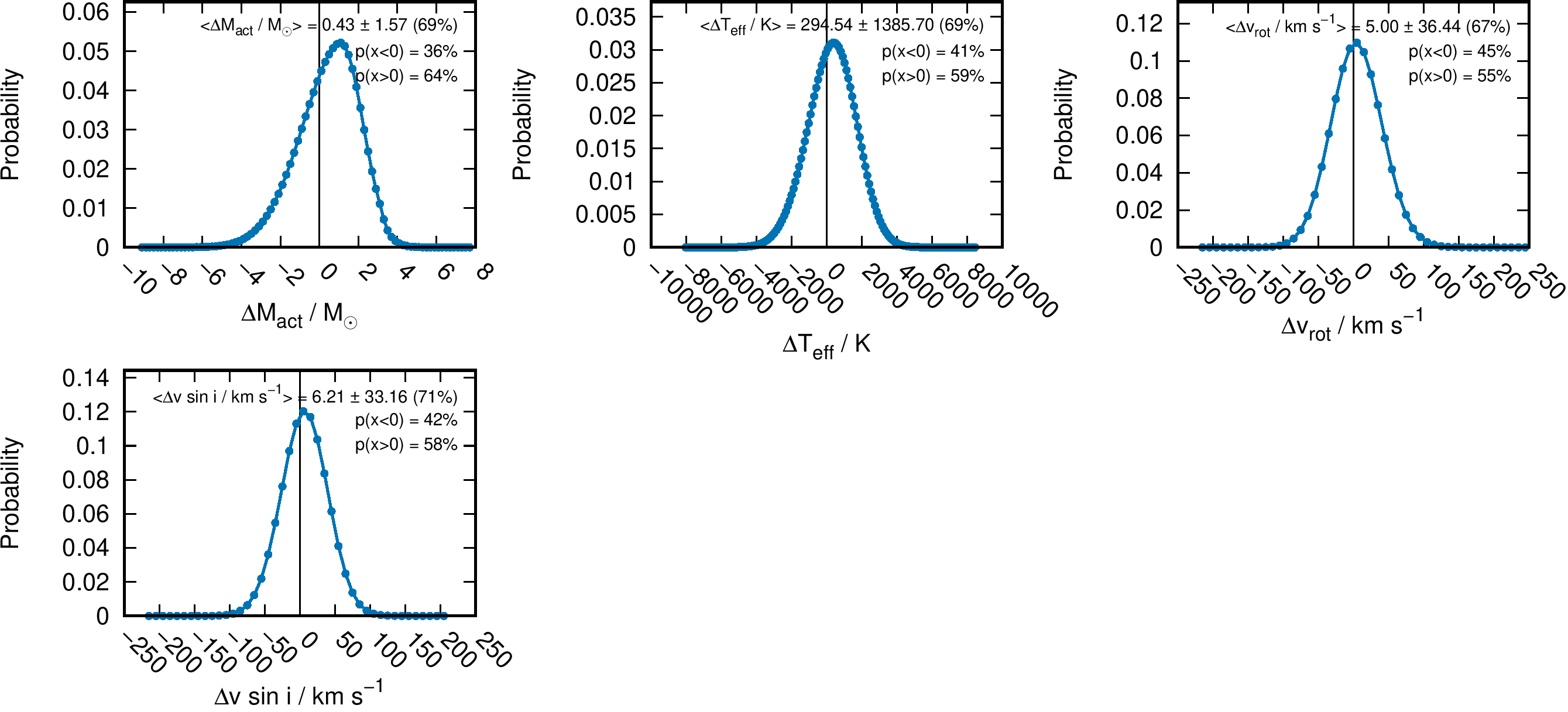}
	\caption{Difference between our prior values and the posterior values of the BONNSAI simulation.}
\label{fig:bonnsai_distribution}
\end{figure*}

As discussed and shown in Sec. \ref{sec:results}, the spectroscopic analysis of the spectrum of S4711 in combination with the photometric information provides properties that can be used for a deeper investigation. We use BONNSAI\footnote{The BONNSAI web-service is available at www.astro.uni-bonn.de/stars/bonnsai.} \citep{Schneider2014} for that. BONNSAI is a Bayesian tool that compares properties of stars with evolution models.
We use a significance level of 5$\%$ with the observables given in Table \ref{tab:bonnsai_input}. The resulting parameters are included in Table \ref{tab:bonnsai_results}. Like the MCMC simulations for the orbital elements of S62, S4711,..., and S4715, the probability distributions of the difference between the prior and posterior values are neglectable small (see Fig. \ref{fig:bonnsai_distribution}).

\begin{table*}[htbp!]
    \centering
    \begin{tabular}{cccc}
            \hline
            \hline
            T$_{\rm eff}$ in [K] & M$_{\rm act}$ in [M$_{\odot}$] & v$\sin{i}$ in [km/s] & v$_{\rm rot}$ in [km/s] \\
            \hline
               11000$^{+1000}_{-1000}$ & 2.20$^{+2.0}_{-1.0}$ & 239.0$^{+25.0}_{-25.0}$ & 263.0$^{+28.0}_{-28.0}$\\ 
             \hline
    \end{tabular}
    \caption{Input parameter of the BONNSAI simulations. The resulting fit parameter can be found in Table \ref{tab:bonnsai_results}. The uncertainties are based on the here presented analysis.}
    \label{tab:bonnsai_input}
\end{table*}

\begin{table*}[htbp!]
    \centering
    \begin{tabular}{ccccccccc}
            \hline
            \hline
            M$_{\rm ini}$ in [M$_{\odot}$] & Age in [Myr] & v$_{\rm ini}$ in [km/s] & log$\frac{L}{L_{\odot}}$ & T$_{\rm eff}$ in [K] & M$_{\rm act}$ in [M$_{\odot}$] & R in [R$_{\odot}$] & log$\frac{g}{cgs}$ & v$_{\rm rot}$ in [km/s] \\
            \hline
               3.20$^{+0.31}_{-0.30}$ & 158.29$^{+60.13}_{-56.15}$ & 280.0$^{+28.84}_{-32.26}$ & 2.15$^{+0.23}_{-0.18}$  & 11587.84$^{+768.39}_{-1128.75}$ & 3.0$^{+0.48}_{-0.10}$ & 2.59$^{+1.71}_{-0.55}$ & 3.92$^{+0.30}_{-0.28}$ & 260.0$^{+24.73}_{-22.58}$ \\ 
             \hline
    \end{tabular}
    \caption{Output parameter of the BONNSAI simulations. The uncertainties are all given at a 1$\sigma$ confidence level. Considering the well defined input parameters, the fit delivers satisfying results.}
    \label{tab:bonnsai_results}
\end{table*}

\section{Data}
\label{sec_app:data_tables}
In this section, we give an overview about the used data for the presented analysis. As mentioned in Sec. \ref{sec:data}, the data was used for the studies in \cite{Parsa2017}, \cite{Peissker2019}, \cite{peissker2020a}, and \cite{peissker2020b}. Partially, the data was also analyzed in \cite{muzic2007}, \cite{muzic2008}, \cite{muzic2010}, \cite{Witzel2012}, \cite{Sabha2012}, \cite{Eckart2013}, \cite{Valencia-S.2015}, and \cite{Shahzamanian2016}. We adapt Table \ref{tab:data_sinfo1}, \ref{tab:naco_data1}, and \ref{tab:naco_data2} from \cite{peissker2020a} and \cite{peissker2020b}.

\begin{table*}[htbp!]
        \centering
        \begin{tabular}{cccccc}
        \hline\hline
        \\      Date & Observation ID  & \multicolumn{3}{c}{Amount of on source exposures} & Exp. Time \\  \cline{3-5} &  & Total & Medium & High &  \\
        (YYYY:MM:DD) &  &  &  &  & (s) \\ \hline\hline 
        
        2007.03.26 & 078.B-0520(A) &  8  &   1  &  2  &    600  \\
        2007.04.22 & 179.B-0261(F) &  7  &   2  &  1  &    600 \\
        2007.04.23 & 179.B-0261(F) &  10 &   0  &  0  &    600 \\
        2007.07.22 & 179.B-0261(F) &  3  &   0  &  2  &    600  \\
        2007.07.24 & 179.B-0261(Z) &  7  &   0  &  7  &    600  \\
        2007.09.03 & 179.B-0261(K) & 11  &   1  &  5  &    600  \\
        2007.09.04 & 179.B-0261(K) &  9  &   0  &  0  &    600  \\   

        \hline  \\
        \end{tabular}   
        \caption{SINFONI data of 2007. The total amount of data is listed. We also list the total amount of data. If the data-cubes do not fulfill a certain quality standard (usually this can be measured by the FHWM of S2), we exclude it from the analysis.}
        \label{tab:data_sinfo1}
        \end{table*}

\begin{table*}[h!]
\centering
\begin{tabular}{ccccc}
\hline
\hline
\multicolumn{5}{c}{NACO}\\
\hline
Date (UT) & Observation ID & \multicolumn{1}{p{1.5cm}}{\centering number \\ of exposures }   & \multicolumn{1}{p{1.5cm}}{\centering Total \\ exposure time(s) } & $\lambda$ \\
\hline
2004-07-06 & 073.B-0775(A) & 344 & 308.04  & K \\
2004-07-08 & 073.B-0775(A) & 285 & 255.82  & K \\
2005-07-25 & 271.B-5019(A) & 330 & 343.76  & K \\
2005-07-27 & 075.B-0093(C) & 158 & 291.09  & K \\
2005-07-29 & 075.B-0093(C) & 101 & 151.74  & K \\
2005-07-30 & 075.B-0093(C) & 187 & 254.07  & K \\
2005-07-30 & 075.B-0093(C) & 266 & 468.50  & K \\
2005-08-02 & 075.B-0093(C) & 80  & 155.77  & K \\
2006-08-02 & 077.B-0014(D) & 48  & 55.36   & K \\
2006-09-23 & 077.B-0014(F) & 48  & 55.15   & K \\
2006-09-24 & 077.B-0014(F) & 53  & 65.10   & K \\
2006-10-03 & 077.B-0014(F) & 48  & 53.84   & K \\
2006-10-20 & 078.B-0136(A) & 47  & 42.79   & K \\
2007-03-04 & 078.B-0136(B) & 48  & 39.86   & K \\
2007-03-20 & 078.B-0136(B) & 96  & 76.19   & K \\
2007-04-04 & 179.B-0261(A) & 63  & 49.87   & K \\ 
2007-05-15 & 079.B-0018(A) & 116 & 181.88  & K \\ 
2008-02-23 & 179.B-0261(L) & 72  & 86.11   & K \\ 
2008-03-13 & 179.B-0261(L) & 96  & 71.49   & K \\ 
2008-04-08 & 179.B-0261(M) & 96  & 71.98   & K \\ 
2009-04-21 & 178.B-0261(W) & 96  & 74.19   & K \\ 
2009-05-03 & 183.B-0100(G) & 144 & 121.73  & K \\ 
2009-05-16 & 183.B-0100(G) & 78  & 82.80   & K \\ 
2009-07-03 & 183.B-0100(D) & 80  & 63.71   & K \\ 
2009-07-04 & 183.B-0100(D) & 80  & 69.72   & K \\ 
2009-07-05 & 183.B-0100(D) & 139 & 110.40  & K \\ 
2009-07-05 & 183.B-0100(D) & 224 & 144.77  & K \\ 
2009-07-06 & 183.B-0100(D) & 56  & 53.81   & K \\ 
2009-07-06 & 183.B-0100(D) & 104 & 72.55   & K \\ 
2009-08-10 & 183.B-0100(I) & 62  & 48.11   & K \\ 
2009-08-12 & 183.B-0100(I) & 101 & 77.32   & K \\
\hline  
\end{tabular}
\caption{NACO data between 2004-2009. The total number of exposures used for the final mosaics is listed. Also, we list the added up total exposure time and the related band.}
\label{tab:naco_data1}
\end{table*}
\begin{table*}[hp!]
\centering
\begin{tabular}{ccccc}
\hline
\hline
\multicolumn{5}{c}{NACO}\\
\hline
Date (UT) & Observation ID & \multicolumn{1}{p{1.5cm}}{\centering number \\ of exposures }   & \multicolumn{1}{p{1.5cm}}{\centering Total \\ exposure time(s) } & $\lambda$ \\
\hline
2010-03-29 & 183.B-0100(L) & 96  & 74.13   & K \\ 
2010-05-09 & 183.B-0100(T) & 12  & 16.63   & K \\ 
2010-05-09 & 183.B-0100(T) & 24  & 42.13   & K \\ 
2010-06-12 & 183.B-0100(T) & 24  & 47.45   & K \\ 
2010-06-16 & 183.B-0100(U) & 48  & 97.78   & K \\
2011-05-27 & 087.B-0017(A) & 305 & 4575    & K \\
2015-08-01 & 095.B-0003(A) & 172 & 5160    & K \\ 
2016-03-22 & 594.B-0498(I) & 144 & 6300    & K \\
\hline  
\end{tabular}
\caption{NACO data between 2010-2016. Please note that NACO was decommissioned between 2013 and 2015.}
\label{tab:naco_data2}
\end{table*}

\bibliography{bib}{}

\begin{thebibliography}{}
\expandafter\ifx\csname natexlab\endcsname\relax\def\natexlab#1{#1}\fi
\providecommand{\url}[1]{\href{#1}{#1}}
\providecommand{\dodoi}[1]{doi:~\href{http://doi.org/#1}{\nolinkurl{#1}}}
\providecommand{\doeprint}[1]{\href{http://ascl.net/#1}{\nolinkurl{http://ascl.net/#1}}}
\providecommand{\doarXiv}[1]{\href{https://arxiv.org/abs/#1}{\nolinkurl{https://arxiv.org/abs/#1}}}

\bibitem[{{Abt} {et~al.}(2002){Abt}, {Levato}, \& {Grosso}}]{Abt2002}
{Abt}, H.~A., {Levato}, H., \& {Grosso}, M. 2002, \apj, 573, 359,
  \dodoi{10.1086/340590}

\bibitem[{{Alexander} \& {Hopman}(2003)}]{Alexander2003b}
{Alexander}, T., \& {Hopman}, C. 2003, \apjl, 590, L29, \dodoi{10.1086/376672}

\bibitem[{{Alexander} \& {Morris}(2003)}]{Alexander2003a}
{Alexander}, T., \& {Morris}, M. 2003, \apjl, 590, L25, \dodoi{10.1086/376671}

\bibitem[{Ali {et~al.}(2020)Ali, Paul, Eckart, Parsa, Zajacek, Pei{\ss}ker,
  Subroweit, Valencia-S., Thomkins, \& Witzel}]{Ali2020}
Ali, B., Paul, D., Eckart, A., {et~al.} 2020, The Astrophysical Journal, 896,
  100, \dodoi{10.3847/1538-4357/ab93ae}

\bibitem[{Almkvist \& Berndt(1988)}]{ramanu1988}
Almkvist, G., \& Berndt, B. 1988, The American Mathematical Monthly, 95, 585.
\newblock \url{http://www.jstor.org/stable/2323302}

\bibitem[{{Bromley} {et~al.}(2012){Bromley}, {Kenyon}, {Geller}, \&
  {Brown}}]{2012ApJ...749L..42B}
{Bromley}, B.~C., {Kenyon}, S.~J., {Geller}, M.~J., \& {Brown}, W.~R. 2012,
  \apjl, 749, L42, \dodoi{10.1088/2041-8205/749/2/L42}

\bibitem[{{Cai} {et~al.}(2018){Cai}, {Liu}, \& {Wang}}]{Cai2018}
{Cai}, R.-G., {Liu}, T.-B., \& {Wang}, S.-J. 2018, arXiv e-prints,
  arXiv:1808.03164.
\newblock \doarXiv{1808.03164}

\bibitem[{{Chen} \& {Amaro-Seoane}(2014)}]{Chen}
{Chen}, X., \& {Amaro-Seoane}, P. 2014, \apjl, 786, L14,
  \dodoi{10.1088/2041-8205/786/2/L14}

\bibitem[{{Clark} {et~al.}(2018){Clark}, {Lohr}, {Najarro}, {Dong}, \&
  {Martins}}]{Clark2018}
{Clark}, J.~S., {Lohr}, M.~E., {Najarro}, F., {Dong}, H., \& {Martins}, F.
  2018, \aap, 617, A65, \dodoi{10.1051/0004-6361/201832826}

\bibitem[{{Clark} \& {Steele}(2000)}]{Clark2000}
{Clark}, J.~S., \& {Steele}, I.~A. 2000, \aaps, 141, 65,
  \dodoi{10.1051/aas:2000310}

\bibitem[{{Do} {et~al.}(2019){Do}, {Witzel}, {Gautam}, {Chen}, {Ghez},
  {Morris}, {Becklin}, {Ciurlo}, {Hosek}, {Martinez}, {Matthews}, {Sakai}, \&
  {Sch{\"o}del}}]{Do2019}
{Do}, T., {Witzel}, G., {Gautam}, A.~K., {et~al.} 2019, \apjl, 882, L27,
  \dodoi{10.3847/2041-8213/ab38c3}

\bibitem[{{Eckart} \& {Genzel}(1996)}]{Eckart1996}
{Eckart}, A., \& {Genzel}, R. 1996, \nat, 383, 415, \dodoi{10.1038/383415a0}

\bibitem[{{Eckart} \& {Genzel}(1997)}]{Eckart1997}
---. 1997, \mnras, 284, 576, \dodoi{10.1093/mnras/284.3.576}

\bibitem[{{Eckart} {et~al.}(2002){Eckart}, {Genzel}, {Ott}, \&
  {Sch{\"o}del}}]{Eckart2002}
{Eckart}, A., {Genzel}, R., {Ott}, T., \& {Sch{\"o}del}, R. 2002, \mnras, 331,
  917, \dodoi{10.1046/j.1365-8711.2002.05237.x}

\bibitem[{{Eckart} {et~al.}(2013){Eckart}, {Mu{\v z}i{\'c}}, {Yazici}, {Sabha},
  {Shahzamanian}, {Witzel}, {Moser}, {Garcia-Marin}, {Valencia-S.}, {Jalali},
  {Bremer}, {Straubmeier}, {Rauch}, {Buchholz}, {Kunneriath}, \&
  {Moultaka}}]{Eckart2013}
{Eckart}, A., {Mu{\v z}i{\'c}}, K., {Yazici}, S., {et~al.} 2013, aap, 551, A18,
  \dodoi{10.1051/0004-6361/201219994}

\bibitem[{{Eisenhauer} {et~al.}(2005){Eisenhauer}, {Genzel}, {Alexander},
  {Abuter}, {Paumard}, {Ott}, {Gilbert}, {Gillessen}, {Horrobin}, {Trippe},
  {Bonnet}, {Dumas}, {Hubin}, {Kaufer}, {Kissler-Patig}, {Monnet},
  {Str{\"o}bele}, {Szeifert}, {Eckart}, {Sch{\"o}del}, \&
  {Zucker}}]{Eisenhauer2005}
{Eisenhauer}, F., {Genzel}, R., {Alexander}, T., {et~al.} 2005, apj, 628, 246,
  \dodoi{10.1086/430667}

\bibitem[{{Foreman-Mackey} {et~al.}(2013){Foreman-Mackey}, {Hogg}, {Lang}, \&
  {Goodman}}]{Foreman-Mackey2013}
{Foreman-Mackey}, D., {Hogg}, D.~W., {Lang}, D., \& {Goodman}, J. 2013, \pasp,
  125, 306, \dodoi{10.1086/670067}

\bibitem[{{Genzel} {et~al.}(2010){Genzel}, {Eisenhauer}, \&
  {Gillessen}}]{2010RvMP...82.3121G}
{Genzel}, R., {Eisenhauer}, F., \& {Gillessen}, S. 2010, Reviews of Modern
  Physics, 82, 3121, \dodoi{10.1103/RevModPhys.82.3121}

\bibitem[{{Ghez} {et~al.}(1998){Ghez}, {Klein}, {Morris}, \&
  {Becklin}}]{Ghez1998}
{Ghez}, A.~M., {Klein}, B.~L., {Morris}, M., \& {Becklin}, E.~E. 1998, \apj,
  509, 678, \dodoi{10.1086/306528}

\bibitem[{{Ghez} {et~al.}(2002){Ghez}, {Duchene}, {Morris}, {Becklin},
  {Hornstein}, {Tanner}, {Kremenek}, {Matthews}, {Thompson}, {Soifer},
  {Larkin}, \& {McLean}}]{Ghez2002}
{Ghez}, A.~M., {Duchene}, G., {Morris}, M., {et~al.} 2002, in American
  Astronomical Society Meeting Abstracts, Vol. 201, 68.04

\bibitem[{{Ghez} {et~al.}(2003){Ghez}, {Duch{\^e}ne}, {Matthews}, {Hornstein},
  {Tanner}, {Larkin}, {Morris}, {Becklin}, {Salim}, {Kremenek}, {Thompson},
  {Soifer}, {Neugebauer}, \& {McLean}}]{Ghez2003}
{Ghez}, A.~M., {Duch{\^e}ne}, G., {Matthews}, K., {et~al.} 2003, \apjl, 586,
  L127, \dodoi{10.1086/374804}

\bibitem[{{Gillessen} {et~al.}(2009){Gillessen}, {Eisenhauer}, {Trippe},
  {Alexander}, {Genzel}, {Martins}, \& {Ott}}]{Gillessen2009}
{Gillessen}, S., {Eisenhauer}, F., {Trippe}, S., {et~al.} 2009, \apj, 692,
  1075, \dodoi{10.1088/0004-637X/692/2/1075}

\bibitem[{{Gillessen} {et~al.}(2017){Gillessen}, {Plewa}, {Eisenhauer}, {Sari},
  {Waisberg}, {Habibi}, {Pfuhl}, {George}, {Dexter}, {von Fellenberg}, {Ott},
  \& {Genzel}}]{Gillessen2017}
{Gillessen}, S., {Plewa}, P.~M., {Eisenhauer}, F., {et~al.} 2017, \apj, 837,
  30, \dodoi{10.3847/1538-4357/aa5c41}

\bibitem[{{Gould} \& {Quillen}(2003)}]{2003ApJ...592..935G}
{Gould}, A., \& {Quillen}, A.~C. 2003, \apj, 592, 935, \dodoi{10.1086/375840}

\bibitem[{{Gravity Collaboration} {et~al.}(2018{\natexlab{a}}){Gravity
  Collaboration}, {Abuter}, {Amorim}, {Anugu}, {Baub{\"o}ck}, {Benisty},
  {Berger}, {Blind}, {Bonnet}, {Brandner}, {Buron}, {Collin}, {Chapron},
  {Cl{\'e}net}, {Coud{\'e} Du Foresto}, {de Zeeuw}, {Deen},
  {Delplancke-Str{\"o}bele}, {Dembet}, {Dexter}, {Duvert}, {Eckart},
  {Eisenhauer}, {Finger}, {F{\"o}rster Schreiber}, {F{\'e}dou}, {Garcia},
  {Garcia Lopez}, {Gao}, {Gendron}, {Genzel}, {Gillessen}, {Gordo}, {Habibi},
  {Haubois}, {Haug}, {Hau{\ss}mann}, {Henning}, {Hippler}, {Horrobin},
  {Hubert}, {Hubin}, {Jimenez Rosales}, {Jochum}, {Jocou}, {Kaufer}, {Kellner},
  {Kendrew}, {Kervella}, {Kok}, {Kulas}, {Lacour}, {Lapeyr{\`e}re}, {Lazareff},
  {Le Bouquin}, {L{\'e}na}, {Lippa}, {Lenzen}, {M{\'e}rand}, {M{\"u}ler},
  {Neumann}, {Ott}, {Palanca}, {Paumard}, {Pasquini}, {Perraut}, {Perrin},
  {Pfuhl}, {Plewa}, {Rabien}, {Ram{\'\i}rez}, {Ramos}, {Rau},
  {Rodr{\'\i}guez-Coira}, {Rohloff}, {Rousset}, {Sanchez-Bermudez},
  {Scheithauer}, {Sch{\"o}ller}, {Schuler}, {Spyromilio}, {Straub},
  {Straubmeier}, {Sturm}, {Tacconi}, {Tristram}, {Vincent}, {von Fellenberg},
  {Wank}, {Waisberg}, {Widmann}, {Wieprecht}, {Wiest}, {Wiezorrek}, {Woillez},
  {Yazici}, {Ziegler}, \& {Zins}}]{gravity2018}
{Gravity Collaboration}, {Abuter}, R., {Amorim}, A., {et~al.}
  2018{\natexlab{a}}, \aap, 615, L15, \dodoi{10.1051/0004-6361/201833718}

\bibitem[{{Gravity Collaboration} {et~al.}(2018{\natexlab{b}}){Gravity
  Collaboration}, {Abuter}, {Amorim}, {Baub{\"o}ck}, {Berger}, {Bonnet}, {Brand
  ner}, {Cl{\'e}net}, {Coud{\'e} Du Foresto}, {de Zeeuw}, {Deen}, {Dexter},
  {Duvert}, {Eckart}, {Eisenhauer}, {F{\"o}rster Schreiber}, {Garcia}, {Gao},
  {Gendron}, {Genzel}, {Gillessen}, {Guajardo}, {Habibi}, {Haubois}, {Henning},
  {Hippler}, {Horrobin}, {Huber}, {Jim{\'e}nez-Rosales}, {Jocou}, {Kervella},
  {Lacour}, {Lapeyr{\`e}re}, {Lazareff}, {Le Bouquin}, {L{\'e}na}, {Lippa},
  {Ott}, {Panduro}, {Paumard}, {Perraut}, {Perrin}, {Pfuhl}, {Plewa}, {Rabien},
  {Rodr{\'\i}guez-Coira}, {Rousset}, {Sternberg}, {Straub}, {Straubmeier},
  {Sturm}, {Tacconi}, {Vincent}, {von Fellenberg}, {Waisberg}, {Widmann},
  {Wieprecht}, {Wiezorrek}, {Woillez}, \& {Yazici}}]{gravity2018b}
---. 2018{\natexlab{b}}, \aap, 618, L10, \dodoi{10.1051/0004-6361/201834294}

\bibitem[{{Gravity Collaboration} {et~al.}(2019){Gravity Collaboration},
  {Abuter}, {Amorim}, {Baub{\"o}ck}, {Berger}, {Bonnet}, {Brand ner},
  {Cl{\'e}net}, {Coud{\'e} Du Foresto}, {de Zeeuw}, {Dexter}, {Duvert},
  {Eckart}, {Eisenhauer}, {F{\"o}rster Schreiber}, {Garcia}, {Gao}, {Gendron},
  {Genzel}, {Gerhard}, {Gillessen}, {Habibi}, {Haubois}, {Henning}, {Hippler},
  {Horrobin}, {Jim{\'e}nez-Rosales}, {Jocou}, {Kervella}, {Lacour},
  {Lapeyr{\`e}re}, {Le Bouquin}, {L{\'e}na}, {Ott}, {Paumard}, {Perraut},
  {Perrin}, {Pfuhl}, {Rabien}, {Rodriguez Coira}, {Rousset}, {Scheithauer},
  {Sternberg}, {Straub}, {Straubmeier}, {Sturm}, {Tacconi}, {Vincent}, {von
  Fellenberg}, {Waisberg}, {Widmann}, {Wieprecht}, {Wiezorrek}, {Woillez}, \&
  {Yazici}}]{Gravity2019}
---. 2019, \aap, 625, L10, \dodoi{10.1051/0004-6361/201935656}

\bibitem[{{Gravity Collaboration} {et~al.}(2020){Gravity Collaboration},
  {Abuter}, {Amorim}, {Baub{\"o}ck}, {Berger}, {Bonnet}, {Brand ner},
  {Cardoso}, {Cl{\'e}net}, {de Zeeuw}, {Dexter}, {Eckart}, {Eisenhauer},
  {F{\"o}rster Schreiber}, {Garcia}, {Gao}, {Gendron}, {Genzel}, {Gillessen},
  {Habibi}, {Haubois}, {Henning}, {Hippler}, {Horrobin}, {Jim{\'e}nez-Rosales},
  {Jochum}, {Jocou}, {Kaufer}, {Kervella}, {Lacour}, {Lapeyr{\`e}re}, {Le
  Bouquin}, {L{\'e}na}, {Nowak}, {Ott}, {Paumard}, {Perraut}, {Perrin},
  {Pfuhl}, {Rodr{\'\i}guez-Coira}, {Shangguan}, {Scheithauer}, {Stadler},
  {Straub}, {Straubmeier}, {Sturm}, {Tacconi}, {Vincent}, {von Fellenberg},
  {Waisberg}, {Widmann}, {Wieprecht}, {Wiezorrek}, {Woillez}, {Yazici}, \&
  {Zins}}]{Gravitycollaboration2020}
---. 2020, \aap, 636, L5, \dodoi{10.1051/0004-6361/202037813}

\bibitem[{{Habibi} {et~al.}(2017){Habibi}, {Gillessen}, {Martins},
  {Eisenhauer}, {Plewa}, {Pfuhl}, {George}, {Dexter}, {Waisberg}, {Ott}, {von
  Fellenberg}, {Baub{\"o}ck}, {Jimenez-Rosales}, \& {Genzel}}]{Habibi2017}
{Habibi}, M., {Gillessen}, S., {Martins}, F., {et~al.} 2017, \apj, 847, 120,
  \dodoi{10.3847/1538-4357/aa876f}

\bibitem[{{Hanson} {et~al.}(1996){Hanson}, {Conti}, \& {Rieke}}]{Hanson1996}
{Hanson}, M.~M., {Conti}, P.~S., \& {Rieke}, M.~J. 1996, \apjs, 107, 281,
  \dodoi{10.1086/192366}

\bibitem[{{Hills}(1988)}]{Hills1988}
{Hills}, J.~G. 1988, \nat, 331, 687, \dodoi{10.1038/331687a0}

\bibitem[{{Hopman} \& {Alexander}(2006)}]{2006ApJ...645.1152H}
{Hopman}, C., \& {Alexander}, T. 2006, \apj, 645, 1152, \dodoi{10.1086/504400}

\bibitem[{{Jalali} {et~al.}(2014){Jalali}, {Pelupessy}, {Eckart}, {Portegies
  Zwart}, {Sabha}, {Borkar}, {Moultaka}, {Mu{\v z}i{\'c}}, \&
  {Moser}}]{Jalali2014}
{Jalali}, B., {Pelupessy}, F.~I., {Eckart}, A., {et~al.} 2014, \mnras, 444,
  1205, \dodoi{10.1093/mnras/stu1483}

\bibitem[{{Jeans}(1919)}]{Jeans}
{Jeans}, J.~H. 1919, \mnras, 79, 408, \dodoi{10.1093/mnras/79.6.408}

\bibitem[{{L{\"o}ckmann} {et~al.}(2008){L{\"o}ckmann}, {Baumgardt}, \&
  {Kroupa}}]{2008ApJ...683L.151L}
{L{\"o}ckmann}, U., {Baumgardt}, H., \& {Kroupa}, P. 2008, \apjl, 683, L151,
  \dodoi{10.1086/591734}

\bibitem[{{Lucy}(1974)}]{Lucy1974}
{Lucy}, L.~B. 1974, \aj, 79, 745, \dodoi{10.1086/111605}

\bibitem[{{Lumsden} {et~al.}(2001){Lumsden}, {Puxley}, \&
  {Hoare}}]{Lumsden2001}
{Lumsden}, S.~L., {Puxley}, P.~J., \& {Hoare}, M.~G. 2001, \mnras, 320, 83,
  \dodoi{10.1046/j.1365-8711.2001.03954.x}

\bibitem[{{Merritt}(2013{\natexlab{a}})}]{2013degn.book.....M}
{Merritt}, D. 2013{\natexlab{a}}, {Dynamics and Evolution of Galactic Nuclei
  (Princeton: Princeton University Press)}

\bibitem[{{Merritt}(2013{\natexlab{b}})}]{Merritt}
---. 2013{\natexlab{b}}, Classical and Quantum Gravity, 30, 244005,
  \dodoi{10.1088/0264-9381/30/24/244005}

\bibitem[{{Meyer} {et~al.}(2006){Meyer}, {Eckart}, {Sch{\"o}del}, {Duschl},
  {Mu{\v{z}}i{\'c}}, {Dov{\v{c}}iak}, \& {Karas}}]{2006A&A...460...15M}
{Meyer}, L., {Eckart}, A., {Sch{\"o}del}, R., {et~al.} 2006, \aap, 460, 15,
  \dodoi{10.1051/0004-6361:20065925}

\bibitem[{{Meyer} {et~al.}(2012){Meyer}, {Ghez}, {Sch{\"o}del}, {Yelda},
  {Boehle}, {Lu}, {Do}, {Morris}, {Becklin}, \& {Matthews}}]{MeyerS552012}
{Meyer}, L., {Ghez}, A.~M., {Sch{\"o}del}, R., {et~al.} 2012, Science, 338, 84,
  \dodoi{10.1126/science.1225506}

\bibitem[{{Mu{\v z}i{\'c}} {et~al.}(2010){Mu{\v z}i{\'c}}, {Eckart},
  {Sch{\"o}del}, {Buchholz}, {Zamaninasab}, \& {Witzel}}]{muzic2010}
{Mu{\v z}i{\'c}}, K., {Eckart}, A., {Sch{\"o}del}, R., {et~al.} 2010, \aap,
  521, A13, \dodoi{10.1051/0004-6361/200913087}

\bibitem[{{Mu{\v z}i{\'c}} {et~al.}(2007){Mu{\v z}i{\'c}}, {Eckart},
  {Sch{\"o}del}, {Meyer}, \& {Zensus}}]{muzic2007}
{Mu{\v z}i{\'c}}, K., {Eckart}, A., {Sch{\"o}del}, R., {Meyer}, L., \&
  {Zensus}, A. 2007, \aap, 469, 993, \dodoi{10.1051/0004-6361:20066265}

\bibitem[{{Mu{\v{z}}i{\'c}} {et~al.}(2008){Mu{\v{z}}i{\'c}}, {Sch{\"o}del},
  {Eckart}, {Meyer}, \& {Zensus}}]{muzic2008}
{Mu{\v{z}}i{\'c}}, K., {Sch{\"o}del}, R., {Eckart}, A., {Meyer}, L., \&
  {Zensus}, A. 2008, \aap, 482, 173, \dodoi{10.1051/0004-6361:20078352}

\bibitem[{{Nayakshin} {et~al.}(2007){Nayakshin}, {Cuadra}, \&
  {Springel}}]{Nayakshin2007}
{Nayakshin}, S., {Cuadra}, J., \& {Springel}, V. 2007, \mnras, 379, 21,
  \dodoi{10.1111/j.1365-2966.2007.11938.x}

\bibitem[{{Parsa} {et~al.}(2017){Parsa}, {Eckart}, {Shahzamanian}, {Karas},
  {Zaja{\v c}ek}, {Zensus}, \& {Straubmeier}}]{Parsa2017}
{Parsa}, M., {Eckart}, A., {Shahzamanian}, B., {et~al.} 2017, \apj, 845, 22,
  \dodoi{10.3847/1538-4357/aa7bf0}

\bibitem[{{Pei{\ss}ker} {et~al.}(2020{\natexlab{a}}){Pei{\ss}ker}, {Eckart}, \&
  {Parsa}}]{peissker2020a}
{Pei{\ss}ker}, F., {Eckart}, A., \& {Parsa}, M. 2020{\natexlab{a}}, \apj, 889,
  61, \dodoi{10.3847/1538-4357/ab5afd}

\bibitem[{{Pei{\ss}ker} {et~al.}(2020{\natexlab{b}}){Pei{\ss}ker}, {Hosseini},
  {Zaja{\v{c}}ek}, {Eckart}, {Saalfeld}, {Valencia-S.}, {Parsa}, \&
  {Karas}}]{peissker2020b}
{Pei{\ss}ker}, F., {Hosseini}, S.~E., {Zaja{\v{c}}ek}, M., {et~al.}
  2020{\natexlab{b}}, \aap, 634, A35, \dodoi{10.1051/0004-6361/201935953}

\bibitem[{{Pei{\ss}ker} {et~al.}(2019){Pei{\ss}ker}, {Zaja{\v{c}}ek}, {Eckart},
  {Sabha}, {Shahzamanian}, \& {Parsa}}]{Peissker2019}
{Pei{\ss}ker}, F., {Zaja{\v{c}}ek}, M., {Eckart}, A., {et~al.} 2019, \aap, 624,
  A97, \dodoi{10.1051/0004-6361/201834947}

\bibitem[{{Perets} {et~al.}(2007{\natexlab{a}}){Perets}, {Hopman}, \&
  {Alexander}}]{2007ApJ...656..709P}
{Perets}, H.~B., {Hopman}, C., \& {Alexander}, T. 2007{\natexlab{a}}, \apj,
  656, 709, \dodoi{10.1086/510377}

\bibitem[{{Perets} {et~al.}(2007{\natexlab{b}}){Perets}, {Hopman}, \&
  {Alexander}}]{Perets}
---. 2007{\natexlab{b}}, \apj, 656, 709, \dodoi{10.1086/510377}

\bibitem[{{Sabha} {et~al.}(2012){Sabha}, {Eckart}, {Merritt}, {Zamaninasab},
  {Witzel}, {Garc{\'{\i}}a-Mar{\'{\i}}n}, {Jalali}, {Valencia-S.}, {Yazici},
  {Buchholz}, {Shahzamanian}, {Rauch}, {Horrobin}, \&
  {Straubmeier}}]{Sabha2012}
{Sabha}, N., {Eckart}, A., {Merritt}, D., {et~al.} 2012, \aap, 545, A70,
  \dodoi{10.1051/0004-6361/201219203}

\bibitem[{{Schneider} {et~al.}(2014){Schneider}, {Langer}, {de Koter}, {Brott},
  {Izzard}, \& {Lau}}]{Schneider2014}
{Schneider}, F.~R.~N., {Langer}, N., {de Koter}, A., {et~al.} 2014, \aap, 570,
  A66, \dodoi{10.1051/0004-6361/201424286}

\bibitem[{{Sch{\"o}del} {et~al.}(2010){Sch{\"o}del}, {Najarro}, {Muzic}, \&
  {Eckart}}]{Schoedel2010}
{Sch{\"o}del}, R., {Najarro}, F., {Muzic}, K., \& {Eckart}, A. 2010, \aap, 511,
  A18, \dodoi{10.1051/0004-6361/200913183}

\bibitem[{{Shahzamanian} {et~al.}(2016){Shahzamanian}, {Eckart}, {Zaja{\v
  c}ek}, {Valencia-S.}, {Sabha}, {Moser}, {Parsa}, {Peissker}, \&
  {Straubmeier}}]{Shahzamanian2016}
{Shahzamanian}, B., {Eckart}, A., {Zaja{\v c}ek}, M., {et~al.} 2016, \aap, 593,
  A131, \dodoi{10.1051/0004-6361/201628994}

\bibitem[{{Slettebak} {et~al.}(1975){Slettebak}, {Collins}, {Boyce}, {White},
  \& {Parkinson}}]{Slettebak1975}
{Slettebak}, A., {Collins}, G.~W., I., {Boyce}, P.~B., {White}, N.~M., \&
  {Parkinson}, T.~D. 1975, \apjs, 29, 137, \dodoi{10.1086/190338}

\bibitem[{{Tokunaga}(2000)}]{Tokunaga2000}
{Tokunaga}, A.~T. 2000, {Infrared Astronomy}, ed. A.~N. {Cox}, 143

\bibitem[{{Valencia-S.} {et~al.}(2015){Valencia-S.}, {Eckart}, {Zaja{\v c}ek},
  {Peissker}, {Parsa}, {Grosso}, {Mossoux}, {Porquet}, {Jalali}, {Karas},
  {Yazici}, {Shahzamanian}, {Sabha}, {Saalfeld}, {Smajic}, {Grellmann},
  {Moser}, {Horrobin}, {Borkar}, {Garc{\'{\i}}a-Mar{\'{\i}}n}, {Dov{\v c}iak},
  {Kunneriath}, {Karssen}, {Bursa}, {Straubmeier}, \&
  {Bushouse}}]{Valencia-S.2015}
{Valencia-S.}, M., {Eckart}, A., {Zaja{\v c}ek}, M., {et~al.} 2015, \apj, 800,
  125, \dodoi{10.1088/0004-637X/800/2/125}

\bibitem[{{Waisberg} {et~al.}(2018){Waisberg}, {Dexter}, {Gillessen}, {Pfuhl},
  {Eisenhauer}, {Plewa}, {Baub{\"o}ck}, {Jimenez-Rosales}, {Habibi}, {Ott},
  {von Fellenberg}, {Gao}, {Widmann}, \& {Genzel}}]{2018MNRAS.476.3600W}
{Waisberg}, I., {Dexter}, J., {Gillessen}, S., {et~al.} 2018, \mnras, 476,
  3600, \dodoi{10.1093/mnras/sty476}

\bibitem[{{Weinberg}(1972)}]{Weinberg1972}
{Weinberg}, S. 1972, {Gravitation and Cosmology: Principles and Applications of
  the General Theory of Relativity}

\bibitem[{{Witzel} {et~al.}(2012){Witzel}, {Eckart}, {Bremer}, {Zamaninasab},
  {Shahzamanian}, {Valencia-S.}, {Sch{\"o}del}, {Karas}, {Lenzen}, {Marchili},
  {Sabha}, {Garcia-Marin}, {Buchholz}, {Kunneriath}, \&
  {Straubmeier}}]{Witzel2012}
{Witzel}, G., {Eckart}, A., {Bremer}, M., {et~al.} 2012, \apjs, 203, 18,
  \dodoi{10.1088/0067-0049/203/2/18}

\bibitem[{{Zaja{\v{c}}ek} {et~al.}(2014){Zaja{\v{c}}ek}, {Karas}, \&
  {Eckart}}]{Zajacek2014}
{Zaja{\v{c}}ek}, M., {Karas}, V., \& {Eckart}, A. 2014, \aap, 565, A17,
  \dodoi{10.1051/0004-6361/201322713}

\bibitem[{{Zaja{\v{c}}ek} \& {Tursunov}(2018)}]{2018AN....339..324Z}
{Zaja{\v{c}}ek}, M., \& {Tursunov}, A.~A. 2018, Astronomische Nachrichten, 339,
  324, \dodoi{10.1002/asna.201813499}

\bibitem[{{Zorec} {et~al.}(2017){Zorec}, {Fr{\'e}mat}, {Domiciano de Souza},
  {Royer}, {Cidale}, {Hubert}, {Semaan}, {Martayan}, {Cochetti}, {Arias},
  {Aidelman}, \& {Stee}}]{Zorec2017}
{Zorec}, J., {Fr{\'e}mat}, Y., {Domiciano de Souza}, A., {et~al.} 2017, \aap,
  602, A83, \dodoi{10.1051/0004-6361/201628761}

\end{thebibliography}
\bibliographystyle{aasjournal}

\end{document}